\documentclass[a4paper,11pt]{article}
\pdfoutput=1
\usepackage{xcolor}
\definecolor{mygreen}{RGB}{0,100,0}
\usepackage{jheppub}
\usepackage[normalem]{ulem}
\usepackage[english]{babel}
\usepackage{amsmath}
\usepackage{amsfonts}
\usepackage{amssymb}
\usepackage{amsthm}
\usepackage{mathtools}
\usepackage{microtype}
\usepackage{enumitem}
\usepackage{float}
\usepackage[noabbrev]{cleveref}
\usepackage{tcolorbox}
\usepackage{subcaption}
\usepackage{wasysym}
\graphicspath{{Images/}}
\usepackage{tikz}
\usetikzlibrary{arrows.meta,decorations.markings}
\definecolor{tealblue}{RGB}{11,94,125}
\definecolor{brickred}{RGB}{164,55,58}
\definecolor{forest}{RGB}{46,125,50}
\newcommand{\cross}[2]{
  \draw[#1, line width=1.15pt, line cap=round]
     ([shift={(-3.3pt,-3.3pt)}]#2) -- ([shift={(3.3pt,3.3pt)}]#2)
     ([shift={(-3.3pt,3.3pt)}]#2) -- ([shift={(3.3pt,-3.3pt)}]#2);
}

\newcommand{\cO}{{\cal O}}

\newcommand{\Gcusp}{\Gamma_{\mathrm{cusp}}}
\newcommand{\Goct}{\Gamma_{\mathrm{oct}}}
\newcommand{\Ghex}{\Gamma_{\mathrm{hex}}}

\makeatletter
\renewcommand*\l@section[2]{%
  \ifnum \c@tocdepth >\z@
    \addpenalty\@secpenalty
    \addvspace{0.45em \@plus\p@}%
    \setlength\@tempdima{1.5em}%
    \begingroup
      \parindent \z@ \rightskip \@pnumwidth
      \parfillskip -\@pnumwidth
      \leavevmode \bfseries
      \advance\leftskip\@tempdima \hskip -\leftskip
      #1\nobreak\hfil \nobreak\hb@xt@\@pnumwidth{\hss #2}\par
    \endgroup
  \fi}
\makeatother
\renewcommand\beforetochook{\pagestyle{myplain}\pagenumbering{roman}\begingroup\small}
\renewcommand\afterTocSpace{\endgroup\medskip}
\preprint{MPP-2026-123}

\title{Positivity properties of observables in planar maximally supersymmetric Yang--Mills theory}

\author[a,b]{Maximilian Haensch}
\affiliation[a]{Max-Planck-Institut f\"ur Physik, Werner-Heisenberg-Institut,\\
Boltzmannstr.~8, 85748 Garching, Germany}
\affiliation[b]{Arnold Sommerfeld Center for Theoretical Physics, Ludwig-Maximilians-Universit\"at M\"unchen,\\
Theresienstr.~37, 80333 M\"unchen, Germany}
\emailAdd{haensch@mpp.mpg.de}

\abstract{
We study positivity properties of exact observables in planar 
$\mathcal{N}=4$ super Yang--Mills as functions of the 't Hooft coupling. Motivated by analogous results in quantum mechanics, we ask whether such observables admit a once-subtracted dispersion representation in the coupling over a positive spectral measure. Our main result is that this property, also known as the \emph{Stieltjes property}, holds for a broad class of exact observables. We prove it analytically, through
integral representations, for the octagon anomalous dimension, the logarithm of the circular
Wilson loop, the Bremsstrahlung function, and anomalous dimensions in the BMN limit,
and we provide numerical evidence for the cusp and tilted cusp anomalous dimensions. We also identify quantities for which the Stieltjes property does not hold, and study the weaker positivity property of complete monotonicity. The Stieltjes property yields two powerful consequences: it lets us turn perturbative input into rigorous non-perturbative bounds, and bootstrap perturbative coefficients. 
We also show how the strong-coupling expansion and its non-perturbative 
corrections can be recovered from the once-subtracted dispersion representation via 
a Mellin--Barnes representation and outline a method to estimate the strong-coupling expansion from weak-coupling data.
}

\begin{document}

\maketitle
\flushbottom
\newpage

\section{Introduction}
\label{sec:intro}

The dynamics of quantum field theories is strongly constrained by
positivity properties that follow from general principles such as
unitarity and locality. A classical example is the
K\"all\'en--Lehmann representation, in which unitarity implies that
two-point functions admit a dispersive representation with a positive
spectral measure in the kinematical variable $p^2$
\cite{Weinberg:1995mt,Eden:1966dnq,Colangelo:2000dp}. Positivity properties have led to
substantial progress in quantum field theory and in the study of scattering amplitudes: low-energy constraints on effective field theories \cite{Adams:2006sv,Bellazzini_2021}, bootstrap programs \cite{Poland:2018epd,Kruczenski:2022lot,ditsch2026approximatingfeynmanintegralsusing}, and positive geometries \cite{Arkani-Hamed:2013jha,henn2025positivitypropertiesscatteringamplitudes,Arkani-Hamed:2020blm}.

Such arguments concern the analytic structure of observables in the
kinematics. By contrast, much less is known about the analytic
structure of observables as functions of the \emph{coupling constant}.
Usually, quantum field theory is only accessible through
perturbation theory, which makes it difficult to investigate analyticity and positivity properties in the coupling dependence of observables. Resurgence provides one means of accessing this structure by reconstructing non-perturbative information from the large-order growth of the perturbative series \cite{Dunne:2015eaa,Aniceto:2018bis,Dorigoni_2019}. This gives insight into the structure of observables around weak coupling, but it is nonetheless difficult to access the global analytic structure of the exact observable and to probe positivity properties. Given how powerful positivity in the kinematic variables has proved to be, it is natural to ask whether analogous properties constrain the coupling dependence, and what they would imply.

Such positivity properties in the coupling are in fact known to hold in quantum mechanics. For example, consider the ground state energy of the quartic anharmonic oscillator with Hamiltonian
\begin{equation}
H=\frac{p^2}{2}+\frac{x^2}{2}+\lambda x^4.
\end{equation}
Bender and Wu \cite{PhysRev.184.1231} showed that the ground state energy $E_0(\lambda)$ can be analytically continued to the complex $\lambda$-plane cut along the negative axis and that it admits a once-subtracted dispersion relation of the form
\begin{equation}\label{eq:QMpositivity}
    \frac{E_0(\lambda)-\frac{1}{2}}{\lambda} = \int_0^\infty dt\, \frac{1}{\lambda+t} \, \mu(t),
\end{equation}
where the spectral density $\mu(t)$ is related to the discontinuity of $E_0$ across the cut. The spectral measure $\mu(t)$ does not depend on $\lambda$. Following this work, Simon \cite{SIMON197076} proved that $\mu(t)$ is positive. This implies that the right-hand side of \eqref{eq:QMpositivity} is a \emph{Stieltjes function} \cite{Bender:1999box}. The perturbative series for $E_0(\lambda)$ is correspondingly a \emph{series of Stieltjes} \cite{Baker_Graves-Morris_1996}. This property is surprisingly powerful. It guarantees that the Pad\'e approximants built from the perturbative series converge to the exact function and give non-perturbative bounds. This is striking because the perturbative expansion for the ground state energy has zero radius of convergence. Although the series diverges for every $\lambda$, positivity alone is enough to reconstruct the exact answer from it. In this sense, Pad\'e approximants are an alternative to Borel resummation usually employed in the context of asymptotic series \cite{Dorigoni_2019,Aniceto:2018bis,Dunne:2015eaa}.
The Stieltjes property has been established in several quantum mechanical models \cite{Bender:2001jq,Graffi:1970erh,Grecchi_2009}. To the best of our knowledge, it has not yet been investigated in quantum field theories, where the coupling dependence of observables is much less understood. 

In planar $\mathcal{N}=4$ super Yang--Mills (sYM), a large number of quantities are known exactly as functions of the 't~Hooft coupling $\lambda = g_{\text{YM}}^2 N$ \cite{Beisert:2006ez,Beisert:2010jr,Pestun_2012,bajnok2024solvingfourdimensionalsuperconformalyangmills}. Moreover, the strong-coupling expansion can be understood from the AdS/CFT correspondence \cite{Maldacena:1997re,Beisert:2006ez}. This makes the theory an ideal testing ground for the coupling dependence of exact observables, and for asking whether positivity properties such as \eqref{eq:QMpositivity} can hold in a quantum field theory.

Motivated by the quantum mechanical example, the main aim of this paper is to investigate the Stieltjes property of exact quantities in planar $\mathcal{N}=4$ sYM. Concretely, we ask whether an observable $\mathcal{O}(\lambda)$ admits a once-subtracted dispersion representation with a positive spectral measure,
\begin{equation}
    \frac{\mathcal{O}(\lambda)}{\lambda} = \int_R^\infty dt \, \frac{1}{\lambda+t}\, \mu(t) \qquad \mu(t) \geq 0,
\end{equation}
where $R$ is the radius of convergence. We then say that $\mathcal{O}(\lambda)/\lambda$ is a Stieltjes function in the coupling. A necessary condition for the Stieltjes property is complete monotonicity, a weaker positivity property that has recently appeared in the study of scattering amplitudes \cite{henn2025positivitypropertiesscatteringamplitudes,ditsch2026approximatingfeynmanintegralsusing}. We develop a method to probe it based on Borel summation. The main result of this paper is that the Stieltjes property holds for a broad class of these exact quantities. We prove it analytically for the octagon anomalous dimension, the logarithm of the circular Wilson loop, anomalous dimensions in the BMN limit, and the Bremsstrahlung function. We provide numerical evidence based on perturbative data for the cusp anomalous dimension and the tilted cusp anomalous dimension. Throughout we work in the planar limit. The circular Wilson loop and the Bremsstrahlung function are also known at finite $N$, and the Stieltjes property holds at finite $N$ as well.

Moreover, we identify two applications of the Stieltjes property. First, it allows us to turn perturbative data into non-perturbative bounds. Second, it allows us to bootstrap perturbative coefficients from the positivity structure. We illustrate both of these applications on the cusp anomalous dimension.
We show how one can recover the strong-coupling expansion, including non-perturbative corrections, from the dispersive representation via a Mellin--Barnes representation, and illustrate this with the octagon anomalous dimension. Using this result, we outline a method to extract strong-coupling coefficients purely from weak-coupling data.

\subsection*{Outline} 
In Section~\ref{sec:stieltjes} we review the
mathematical definitions of Stieltjes
functions and the Hankel matrix characterisation of the associated
moment problem. In Section~\ref{sec:observables} we collect exact
representations of observables in planar $\mathcal{N}=4$ sYM. The main result of the paper
is Section~\ref{sec:positivity}, where we investigate the Stieltjes property of exact observables. We give analytic proofs based on integral representation where available and provide numerical tests based on perturbative expansions. We also comment on observables for which the Stieltjes property does not hold. In Section~\ref{sec:applications}
we develop two applications of the Stieltjes property: Pad\'e bounds and a
bootstrap for perturbative coefficients. We apply these methods to the cusp anomalous dimension. In Section~\ref{sec:strong} we show how one can recover the strong-coupling expansion from the dispersive
representation and illustrate this with the
octagon anomalous dimension. Moreover, we outline a method to extract strong-coupling expansion coefficients from weak-coupling data, and illustrate this method on the cusp anomalous dimension. 
In Section~\ref{sec:conclusion} we summarize our findings and comment on possible future directions. In Appendix~\ref{app:CM}, we investigate complete monotonicity. We develop a method to prove complete monotonicity based on Borel summation and an asymptotic estimate from strong-coupling information. Moreover, we illustrate this on the cusp anomalous dimension and the octagon correlation function. In
Appendix~\ref{app:circular}, we show how the strong-coupling expansion of the logarithm of the circular Wilson loop can be computed from its dispersive representation.

\section{Stieltjes functions and moment problems}
\label{sec:stieltjes}

We collect the definitions and results that we shall need. A full
treatment of Stieltjes functions, series of Stieltjes, and completely monotonic functions may be found in
\cite{WidderWidder+2015,Baker_Graves-Morris_1996,Bender:1999box,schmudgen2020lecturesmomentproblem,merkle2012completelymonotonefunctions}.
Our conventions follow
\cite{henn2025positivitypropertiesscatteringamplitudes,ditsch2026approximatingfeynmanintegralsusing}.

\subsection{Stieltjes functions}
\label{sec:stieltjes-def}

Let $f:\mathbb{C}\setminus(-\infty,-R]\to\mathbb{C}$, $R>0$, be an analytic function in
the cut plane that goes to zero as $|z|\to\infty$. Cauchy's theorem and a contour deformation lead to the dispersive representation 
\begin{equation}
f(z)=\int_R^\infty dt\, \frac{\mu(t)}{z+t}, \qquad \mu(t) = -\frac{1}{2\pi i} \mathrm{disc}f(-t).
\label{eq:stieltjes-rep}
\end{equation}
Here $f(z)$ is called a \emph{Stieltjes function} if the spectral density $\mu(t)$ is positive for all $t\ge R$.
Krein's theorem
\cite{doi:10.1137/1.9781611976397.ch3,Raman_2021} conversely
states that $f$ is a Stieltjes function if and only if $f(x)\ge 0$ for $x>0$ and $f$ admits
an analytic continuation to $\mathbb{C}\setminus(-\infty,-R]$ such that $\operatorname{Im}f(z)\leq0$ for $\operatorname{Im}z\geq0$. 

In practice, it is often easier to work with the equivalent representation
\begin{equation}
f(z)
=\int_0^{1/R} dt\, \frac{\nu(t)}{1+t z},
\qquad \nu(t)=\mu(1/t)/t\ge 0,
\label{eq:stieltjes-rep-nu}
\end{equation}
obtained by the change of variables $t\mapsto 1/t$.

In the following, we consider observables that grow like $\sqrt{\lambda}$ at strong coupling. These functions do not admit
\eqref{eq:stieltjes-rep} directly, but may admit a \emph{once-subtracted}
dispersion relation
\begin{equation}
\frac{\mathcal{O}(\lambda)}{\lambda}
=\int_0^{1/R} dt\, \frac{\nu(t)}{1+t\lambda}.
\label{eq:once-subtracted}
\end{equation}
Here we assumed that $\mathcal{O}(\lambda)$ is normalized such that $\mathcal{O}(\lambda)/\lambda$ is finite as $\lambda \to 0$.
In the case that $\nu(t)\geq0$, $\mathcal{O}(\lambda)/\lambda$ defines a Stieltjes function.
The additional factor of~$\lambda$ is inessential for the main results below.
By slight abuse of terminology, we will refer to any $\mathcal{O}$ obeying
\eqref{eq:once-subtracted} as Stieltjes.

It is difficult to prove positivity analytically, since this requires an understanding of the analytic structure of the observables in the complex coupling plane. However, the Stieltjes property is equivalent to a set of conditions on the perturbative coefficients, which we explain in the following subsection. 

\subsection{Series of Stieltjes and moment problems}
\label{sec:moment-problems}

Expanding \eqref{eq:stieltjes-rep-nu} in powers of $z$ gives a power
series\footnote{For $R>0$ the exchange of summation and integration is justified. For $R=0$ it gives rise to an asymptotic series. However, this case is not relevant for our applications.}
\begin{equation}
f(z)=\sum_{n=0}^\infty (-1)^n a_n z^n,
\qquad
a_n=\int_0^{1/R} dt\, t^n\nu(t).
\label{eq:series-of-stieltjes}
\end{equation}
The coefficients $a_n$ are thus the moments of the positive measure
$\nu(t)$ supported on $[0,1/R]$: they are the solution of a
\emph{Hausdorff moment problem} \cite{doi:10.1137/1.9781611976397.ch3,schmudgen2020lecturesmomentproblem}. Series with this property are also known as \emph{series of Stieltjes}. Sharp necessary and sufficient
conditions for a sequence $(a_n)$ to arise from such a measure were recently
worked out in \cite{Bellazzini_2021}. The conditions are most conveniently expressed in terms of four Hankel matrices
$H_M^N=\{a_{i+j+M}\}_{i,j=0}^{\lfloor(N-M)/2\rfloor}$, and read
\begin{equation}
H_0^N\succeq 0,\qquad  H_1^N\succeq 0,\qquad
\frac{1}{R}\,H_0^{N-1}-H_1^N\succeq 0,\qquad
\frac{1}{R}\,H_1^{N-1}-H_2^N\succeq 0, \quad \forall N>0.
\label{eq:hankel-constraints}
\end{equation}
Here, $\succeq 0$ denotes positive semidefiniteness, which is equivalent to having only non-negative eigenvalues. 
For a convergent series
($R>0$), conditions \eqref{eq:hankel-constraints} are equivalent to
\eqref{eq:stieltjes-rep-nu} with $\nu\ge0$. This is the key
relation that lets us test the Stieltjes property purely from
perturbation theory. 

For a once-subtracted dispersion relation, the same conditions hold with $a_n$ replaced by $a_{n+1}$.

\begin{tcolorbox}
For an observable $\mathcal{O}(\lambda)$ analytic at the origin, the
once-subtracted dispersion relation \eqref{eq:once-subtracted} with
$\nu\ge 0$ holds if and only if the perturbative coefficients
$a_{n+1}$ satisfy the Hankel
constraints \eqref{eq:hankel-constraints}. \noindent
\end{tcolorbox} \noindent
For example, truncating the constraints at $N=3$ yields the necessary conditions
\begin{equation}
a_1\ge 0,\qquad a_1 a_3\ge a_2^2,\qquad
a_2\ge 0,\qquad
\frac{a_1}{R}\ge a_2,\qquad
\frac{a_2}{R}\ge a_3. \label{eq:HankelEx}
\end{equation}
In practice truncating the Hankel matrix constraints at finite $N$ provides a useful test for the Stieltjes property. A violation rules out the possibility that an observable is Stieltjes, whereas constraints that remain satisfied to high orders provide strong evidence in its favor.
\subsection{Pad\'e approximants}
\label{sec:pade}
Pad\'e approximants provide the best rational approximation to a function $f(x)$. The Pad\'e approximant $P^M_N(x)$ of order $[M,N]$ is defined as the ratio of two polynomials of degree $M$ and $N$ respectively, such that the Taylor expansion of $P^M_N$ around $x=0$ agrees with that of $f$ up to order $M+N$ \cite{Bender:1999box,Baker_Graves-Morris_1996}.
These approximants are commonly used in physics applications, since they often have better convergence properties than the original Taylor series, and can be used to continue a function beyond the radius of convergence of its Taylor expansion \cite{Basdevant:1972fe}. However, in general, Pad\'e approximants are a heuristic tool, and their convergence properties are not well understood.

Positivity controls the convergence of Pad\'e approximants and allows us to make rigorous statements. Given a
series of Stieltjes \eqref{eq:series-of-stieltjes} with $R>0$, the
diagonal and near-diagonal Pad\'e approximants $P^N_N$ and
$P^{N+1}_N$ have the following properties
\cite{Baker_Graves-Morris_1996,Bender:1999box,ditsch2026approximatingfeynmanintegralsusing}:
\begin{enumerate}[label=(\roman*)]
\item all poles lie on the real axis $(-\infty,-R)$ and all residues
are positive.
\item for $x\in(-R,\infty)$, the sequence $(P^N_N(x))$ decreases and
$(P^{N}_{N-1}(x))$ increases with $N$, and
\begin{equation}
P^N_N(x)\ \ge\ f(x)\ \ge\ P^{N}_{N-1}(x),
\qquad x>-R;
\end{equation}
\item both sequences converge pointwise to $f$ on
$\mathbb{C}\setminus(-\infty,-R]$.
\end{enumerate}
For a once-subtracted Stieltjes function, one has the same properties with 
$P^{N+1}_N(\lambda)\ge \mathcal{O}(\lambda)\ge P^N_N(\lambda)$.

Positivity therefore turns Pad\'e approximants built from perturbative data into rigorous,
two-sided, non-perturbative bounds.

\section{Exact observables in planar $\mathcal{N}=4$ sYM}
\label{sec:observables}

Planar $\mathcal{N}=4$ sYM offers a unique setting in which to investigate the coupling dependence of observables. Progress in integrability and localization has made it possible to compute many observables exactly as functions of the 't~Hooft coupling~$\lambda$, either through closed-form expressions or through integral representations that can be evaluated to any desired order in perturbation theory. This wealth of exact data allows us to test whether the positivity structures found in quantum mechanical models persist in quantum field theory. 
In this section we review the exact observables we shall analyze for positivity. Throughout, $g^2=\lambda/(4\pi)^2$, where $g$ denotes the planar Yang--Mills coupling, $\lambda=g_{\text{YM}}^2 N$ the 't~Hooft coupling, and $g_{\text{YM}}$ the Yang--Mills coupling as it appears in the Lagrangian. 

\subsection{Cusp anomalous dimension}

The cusp anomalous dimension $\Gcusp$ governs the UV divergence of a
light-like Wilson loop with a cusp
\cite{Polyakov:1980ca,Korchemsky:1987wg}. It also controls the large
spin limit of twist-two anomalous dimensions and the soft/collinear
divergences of amplitudes
\cite{KORCHEMSKAYA1992169,collins2004factorizationhardprocessesqcd,Magnea:1990zb,Kodaira:1981nh,Alday:2007mf,Korchemsky:1988si}.
In planar $\mathcal{N}=4$ sYM, it can be computed at finite coupling via the BES equation
\cite{Beisert:2006ez}, which can be rewritten as an
expression involving a semi-infinite matrix of Bessel-function
integrals \cite{Benna:2006nd}. With
\begin{equation}
K_{ij}=2j(-1)^{j(i+1)}\int_0^\infty \frac{dt}{t}
\frac{J_i(2gt)J_j(2gt)}{e^t-1},
\label{eq:Kij}
\end{equation}
split into odd-odd, odd-even, even-odd, and even-even blocks $K_{\circ\circ}, K_{\circ\bullet}, K_{\bullet\circ}, K_{\bullet\bullet}$,
the semi-infinite matrix
$K=\bigl(\begin{smallmatrix}K_{\circ\circ}&K_{\circ\bullet}\\ K_{\bullet\circ}&K_{\bullet\bullet}\end{smallmatrix}\bigr)$
gives
\begin{equation}
\Gcusp=4g^2\left(\frac{1}{1+K}\right)_{11}.
\label{eq:Gcusp}
\end{equation}
Expanding $(1+K)^{-1}=\sum_{n\ge0}(-K)^n$ produces the weak-coupling
series,
\begin{equation}
\Gcusp=4g^2-\tfrac{4}{3}\pi^2 g^4+\tfrac{44}{45}\pi^4 g^6
-\bigl(\tfrac{292}{315}\pi^6+32\,\zeta(3)^2\bigr)g^8+\cdots,
\label{eq:Gcusp-weak}
\end{equation}
which may be computed to very high loop order. The
strong-coupling expansion reads
\cite{Basso_2008}
\begin{equation}
\Gcusp(\lambda)\sim
\frac{\sqrt{\lambda}}{2\pi}-\frac{3\log2}{2\pi}-\frac{G}{2\pi\sqrt\lambda}
+\cdots,\qquad \lambda\to\infty,
\label{eq:Gcusp-strong}
\end{equation}
with $G$ the Catalan constant. Its resurgence structure was worked out
in \cite{Dorigoni_2015,Aniceto:2015rua}.

\subsection{Tilted cusp anomalous dimension}
A one-parameter generalisation $\Gamma(\alpha)$ is obtained by
``tilting'' the BES matrix \cite{Basso_2020}:
\begin{equation}
K(\alpha)=2\cos\alpha
\begin{pmatrix}\cos\alpha\, K_{\circ\circ} & \sin\alpha\, K_{\circ\bullet}\\
\sin\alpha\, K_{\bullet\circ} & \cos\alpha\, K_{\bullet\bullet}\end{pmatrix},
\qquad
\Gamma(\alpha)=4g^2\left(\frac{1}{1+K(\alpha)}\right)_{11}.
\label{eq:tilted}
\end{equation}
At $\alpha=0$ and $\alpha=\pi/4$ it reproduces the octagon anomalous dimension $\Goct$ and the cusp anomalous dimension $\Gcusp$ respectively. At $\alpha = \pi/3$ it defines the hexagon anomalous dimension $\Ghex$. All three govern the leading logarithmic behaviour of
the six-gluon MHV amplitude in the regime where the three conformal
cross-ratios tend to zero \cite{Basso_2020}. Other values of $\alpha$ also appear in higher-point MHV amplitudes \cite{Basso:2022ruw} and in null limits of large-charge correlators \cite{Basso:2026vmx}. The octagon anomalous dimension $\Goct$ also appears in large-charge correlators \cite{Caron-Huot:2021usw,Kostov:2019auq,Belitsky:2019fan} and in $\operatorname{tr} \phi^3$ form factors \cite{Basso:2022ruw}.

\subsection{Bremsstrahlung function and circular Wilson loop}
The angle-dependent cusp anomalous dimension $\Gcusp(\phi)$ governs the UV divergence of a Wilson loop with a cusp of Euclidean angle $\phi$. 
At small angle $\phi$, it defines the Bremsstrahlung function $B$ \cite{Correa_2012}, $\Gcusp(\phi)\sim B(\lambda)\phi^2$, which can be computed exactly for any rank of the gauge group
\begin{equation}
B(\lambda)=\frac{1}{2\pi^2}\lambda \partial_\lambda\log W(\lambda),\qquad
W(\lambda)=\langle W_{\mathrm{circ}}\rangle.
\end{equation}
Here, $\langle W_{\mathrm{circ}} \rangle $ is the expectation value of the $\tfrac{1}{2}$-BPS circular Wilson loop, which can be computed at finite $N$ using localization \cite{Erickson_2000,Drukker_2001,Pestun_2012}
\begin{equation}\label{eq:W-circular}
    W(\lambda) =\frac{1}{N}L^1_{N-1}(-\lambda/4N)e^{\lambda/8N}  \quad \underset{N \to \infty}{\longrightarrow} \quad \frac{2}{\sqrt\lambda}I_1(\sqrt\lambda).
\end{equation}
$L^1_N$ are the associated Laguerre polynomials.

\subsection{A class of observables described by Fredholm determinants}
\label{sec:fredholm}

A striking recent development is that many exact observables in planar
$\mathcal{N}=4$ sYM can be written as Fredholm determinants of a
semi-infinite matrix of Bessel-function integrals
\cite{Belitsky_2020,Belitsky_2021,Bajnok:2024epf,Bajnok:2024bqr,bajnok2024solvingfourdimensionalsuperconformalyangmills}.
Define
\begin{align} \label{eq:Dl}
D_\ell(g)&=\det\bigl(\delta_{nm}-K_{nm}(g)\bigr)_{n,m\ge 1},\nonumber\\
K_{nm}(g)&=\sqrt{(2n+\ell-1)(2m+\ell-1)}\,(-1)^{n+m}
\int_0^\infty\! \frac{dx}{x}\, J_{2n+\ell-1}(\sqrt x)J_{2m+\ell-1}(\sqrt x)\,\chi\!\left(\tfrac{\sqrt x}{2g}\right),
\end{align}
where $\chi$ is called the symbol function and $\ell\in\mathbb{Z}_{\ge 0}$. Note that depending on $\ell$,
either only even or only odd Bessel functions contribute. This differs structurally from the (tilted) cusp anomalous dimension, where both even and odd indices contribute. Different
choices of $\chi$ yield different physical observables:
\begin{itemize}[leftmargin=*]
\item $\chi_{\mathrm{f.t.}}(x)=2/(1-e^x)$, which describes flux tube correlators \cite{Korchemsky:2025mla,Beccaria:2022ypy,Bajnok:2024epf}. $\log D_\ell^{\mathrm{f.t.}}$
for $\ell=0,1,2$ are known in closed form (cf. equation (2.9) in
\cite{Korchemsky:2025mla}). For $\ell = 0$ one finds
\begin{equation} \label{eq:ftD0}
    \log D^{\mathrm{f.t.}}_0 = \frac{3}{8} \log \Big[ \cosh \left(\frac{\sqrt{\lambda}}{2}\right)\Big]- \frac{1}{8}\log \Big[\frac{\sinh(\tfrac{\sqrt{\lambda}}2)}{\sqrt{\lambda}/2}\Big]. 
\end{equation}
\item $\chi_W(x)=-(2\pi)^2/x^2$, which describes the $\tfrac12$-BPS circular Wilson
loop for $\ell=2$: $\log D_2^{W}=\log W$.
\item 
The octagon describes a specific four-point function of $\tfrac12$-BPS operators in the limit where the scaling dimensions $\Delta$ go to infinity \cite{Chicherin_2016,Coronado_2020,Belitsky_2020,Belitsky_2021},
\begin{equation}
\langle \mathcal{O}_1(x_1)\mathcal{O}_2(x_2)\mathcal{O}_1(x_3)\mathcal{O}_3(x_4)\rangle = \frac{\mathbb{O}(z,\bar{z})^2}{(x_{12}^2 x_{34}^2 x_{13}^2 x_{24}^2)^{\Delta/2}}.
\end{equation}
Conformal symmetry restricts the numerator to depend on two conformal cross-ratios,
\begin{equation}
    \frac{x_{12}^2 x_{34}^2}{x_{13}^2 x_{24}^2} = z\bar{z}, \qquad \frac{x_{23}^2x_{41}^2}{x_{13}^2 x_{24}^2}=(1-z)(1-\bar{z}), \qquad x_{ij}^2=(x_i-x_j)^2.
\end{equation}
In Euclidean kinematics $z$ is complex and satisfies $z^* = \bar{z}$, whereas in Lorentzian kinematics $z$ and $\bar{z}$ are real and independent. The octagon $\mathbb{O}(z,\bar{z})$ admits a representation as a Fredholm determinant. To write it, we introduce parameters $y$ and $\xi$,
\begin{equation}
    z = - e^{-\xi -y}, \qquad \bar{z} = - e^{-\xi +y}.
\end{equation}
In Euclidean kinematics $y$ is purely imaginary, with $iy \in [0,\pi]$ playing the role of an angular variable, and $\xi$ is real. When $y$ and $\xi$ are both real the configuration is Lorentzian. The symbol function reads
\begin{equation}
\chi_{\mathrm{oct}}(x\,|\,y,\xi)=\frac{\cosh y+\cosh\xi}{\cosh y+\cosh\sqrt{x^2+\xi^2}}.
\end{equation}
In the Lorentzian light-like limit of the correlation function, corresponding to fixed $\xi$ and $y\to\infty$, the octagon simplifies to \cite{Belitsky:2019fan}
\begin{align}\label{eq:Goct-closed}
    \log D_0&\sim - y^2 \frac{\Goct(\lambda)}{2\pi^2} - \frac{1}{8}C(\lambda) + g^2 \xi^2 + \mathcal{O}\big( e^{-y} \big) \\
\Goct(\lambda)&=\log\cosh(\sqrt\lambda/2), \\
C(\lambda)&=\log \frac{\sinh (\sqrt \lambda /2)}{\sqrt{\lambda}/2}, \label{eq:clambda}
\end{align}
where $\Goct$ is the octagon anomalous dimension, which also appears in the tilted cusp anomalous dimension \eqref{eq:tilted} at $\alpha=0$. Beyond the light-like limit, the defining equations for the octagon also simplify considerably at the kinematic point $y=\xi=0$, which corresponds to $z=\bar{z}=-1$.

The integer $\ell$ is related to the bridge length of the octagon \cite{Belitsky_2021}. We will focus on the octagon with zero bridge length in the following.
\end{itemize}
The weak-coupling expansion of $\log D_\ell$ is obtained by expanding
the Bessel functions in the integrand, at which point the matrix
elements depend on the symbol only through its moments
$q_n=\int_0^\infty dx\, x^n\chi(x)$. For $\ell=0$,
\begin{equation}
\log D_0=-2g^2 q_1+2g^4(q_3-q_1^2)-g^6(\tfrac{8}{3}q_1^3-4q_1q_3-q_5)+\cdots.
\label{eq:logD0-weak}
\end{equation}
The radius of convergence of the weak-coupling expansion can be computed from the symbol function.
The strong-coupling expansion is controlled by the
Szeg\H{o}--Akhiezer--Kac formula
\cite{Belitsky_2020,Belitsky_2021,Bajnok:2024epf,Bajnok:2024bqr,bajnok2024solvingfourdimensionalsuperconformalyangmills},
\begin{equation}
\log D_\ell\sim
-\frac{A_0}{4\pi}\sqrt\lambda+\frac{1}{2}A_1\log\sqrt\lambda+B
+\cO(1/\sqrt\lambda),\qquad \lambda\to\infty,
\label{eq:SAK}
\end{equation}
with $A_0,A_1,B$ computable from $\chi$ and $\ell$. The subleading
perturbative coefficients are also accessible, and the series is
generically Borel non-summable, which indicates non-perturbative corrections of order
$\cO(e^{-\sqrt\lambda})$
\cite{Belitsky_2021,bajnok2025universalityresurgencegeneralizedtracywidom}. This is like in the case of the cusp anomalous dimension \cite{Basso_2008,Dorigoni_2015}.

\subsection{Konishi anomalous dimension and the BMN limit}
The Konishi operator is given by $\mathcal{K}=\operatorname{tr} \phi^i \phi^i$, where $\phi^i$ denote the six real scalars in $\mathcal{N}=4$ sYM. We will be interested in its anomalous dimension $\Delta_K$. This example differs from the other observables discussed in this section in that no exact representation for $\Delta_K$ is known. Nevertheless, it can be computed to high orders in perturbation theory \cite{Marboe_2015}.

The computation of $\Delta_K$ corresponds to a problem in a spin chain with $L=2$ sites \cite{Minahan:2002ve}. The contributions to $\Delta_K$ divide into a piece captured by the asymptotic Bethe ansatz and a piece from ``wrapping'' corrections \cite{Ambjorn:2005wa,Bajnok:2008bm}. The former can be computed to arbitrary loop order \cite{Beisert:2005fw}. We will discuss these pieces separately in Section~\ref{sec:positivity}.

A complementary regime is provided by the BMN limit \cite{Berenstein:2002jq}. Here one considers BMN operators built from a large number $J$ of complex scalars $Z=\phi^5 + i \phi^6$ with a few impurities \cite{Plefka:2005bk}. As $J\to\infty$ the length of the operator grows, so the wrapping corrections mentioned above drop out. The anomalous dimension then becomes a function of the effective coupling $\lambda' = \lambda/J^2$. Writing $N_n$ for the occupation number for string oscillator mode $n$, one has
\begin{equation}\label{eq:BMNdimensions}
    \Delta - J = \sum_{n=-\infty}^\infty N_n\sqrt{1+n^2 \lambda'}.
\end{equation}

\subsection{Common analytic features}
We briefly collect structural features shared by the exact quantities above.
For all these quantities, the perturbative expansion is sign-alternating in $\lambda$, has a finite radius of convergence $R>0$ (see also the discussion in \cite{Arkani_Hamed_2022}), and the observable grows like $\sqrt\lambda$ at infinity. The last point holds for all observables except the Konishi anomalous dimension, which instead grows like $\lambda^{1/4}$ at strong coupling \cite{Roiban:2011fe}.

These features are consistent with the Stieltjes property. The sign-alternating perturbative expansion indicates a branch point on the negative real axis, while the behavior at strong coupling is consistent with the once-subtracted dispersion relation \eqref{eq:once-subtracted}. Moreover, the radius of convergence is known and the integrability representation can be easily expanded to high orders at weak coupling, making it straightforward to check the Hankel matrix constraints \eqref{eq:hankel-constraints}.

\section{Positivity results}
\label{sec:positivity}

We investigate whether the observables defined in Section~\ref{sec:observables} satisfy a similar positivity property to that of the ground state energy of the anharmonic oscillator \eqref{eq:QMpositivity}.
That is, we ask whether the observable $\mathcal{O}(\lambda)$ admits a once-subtracted dispersion relation of the form \eqref{eq:once-subtracted} with a positive spectral measure $\nu(t)\ge 0$. 

To establish the dispersive representation, it is necessary to understand the analytic structure of the observables as a function of the coupling. However, the defining equations such as \eqref{eq:Gcusp} and \eqref{eq:Dl} are difficult to analyze for complex $\lambda$, since the matrix elements $K_{nm}$ are defined by integrals, which do not converge everywhere in the complex plane.
To establish the existence of a dispersive representation, we need to carefully analytically continue the matrix elements and show that the resulting infinite-dimensional determinant has the right analytic structure.
This is a very subtle problem, which we have not been able to solve. However, we can bypass this problem. In cases where an explicit form is known, like for the octagon anomalous dimension and circular Wilson loop, the analytic continuation is easy, and the dispersive representation can be investigated directly. Furthermore, in the remaining cases, we can exploit that the perturbative expansion has a finite radius of convergence and relate positivity to the moment problem constraints \eqref{eq:hankel-constraints}, which we can check to high loop orders thanks to the matrix representations. 
\subsection{Analytic Stieltjes representations}
\label{sec:analytic-proofs}
In this subsection, we establish the Stieltjes property for observables that admit a closed-form expression, which allows us to find an explicit dispersive representation.
\subsubsection{Octagon anomalous dimension} \label{sec:octagonanomaloussec}
The octagon anomalous dimension admits the closed form expression \cite{Belitsky:2019fan}
\begin{equation}
\Goct(\lambda)= \log \cosh\Big(\frac{\sqrt\lambda}{2}\Big).
\end{equation}
Notice that $\exp \left(\Goct(\lambda)\right)$ is an entire function, and we can use the Weierstrass factorisation theorem to express it as an infinite product over its zeros, which are located at $\lambda_n = -4 \pi^2 (n+\tfrac12)^2 $. In particular, we have
\begin{align}
    \Goct(\lambda) &= \log \prod_{n=0}^\infty \left(1 + \frac{\lambda}{4\pi^2(n+\tfrac12)^2}\right)\\
    &= \sum_{n=0}^\infty \log\!\left(1+\frac{\lambda}{4\pi^2(n+\tfrac12)^2}\right)\\
&=\int_0^{\tfrac{1}{\pi^2}} dt\,\frac{\lambda}{1+t\lambda}\,\sum_{n=0}^\infty\theta\!\left(\frac{1}{4\pi^2(n+\tfrac12)^2}-t\right),
\label{eq:Goct-Stieltjes}
\end{align}
where in the last step we used that 
\begin{equation}
    \log\left(1+\frac{\lambda}{A}\right) = \int_0^\infty dt \, \frac{\lambda}{1+\lambda t} \theta\left( \frac1A -t\right), \quad A>0.
\end{equation}
Equation \eqref{eq:Goct-Stieltjes} is a spectral representation of the form \eqref{eq:once-subtracted} with $R=\pi^2$. The spectral measure is given by a sum over step functions, which is positive for all $t$. 
The first branch point occurs at $\lambda=-\pi^2$, which sets the radius of convergence of the perturbative expansion. 

With an identical argument applied to the function $C(\lambda)$ in \eqref{eq:clambda}, using its factorization over the zeros at $\lambda_n=-4\pi^2 (n+1)^2$,
one can prove the Stieltjes property for $C(\lambda)$. Thus, both functions controlling the null limit of the octagon \eqref{eq:Goct-closed} are Stieltjes functions. 

\subsubsection{Circular Wilson loop}
We can apply the same type of argument to the circular Wilson loop, which in the planar limit is given by
\begin{equation}
    W(\lambda)=(2/\sqrt\lambda) I_1(\sqrt\lambda).
\end{equation}
This is an entire function with zeros located at $\lambda_k=-j_{1,k}^2$, where $j_{1,k}$ is the $k$th positive zero of the Bessel function $J_1$. Using Weierstrass factorisation, we can express the circular Wilson loop as an infinite product. We will be interested in studying the logarithm of the circular Wilson loop, which can be expressed as
\begin{align}
    \log W(\lambda) &= \sum_{k=1}^\infty \log\!\left(1+\frac{\lambda}{j_{1,k}^2}\right) \\
    &=\int_0^{\tfrac{1}{j_{1,1}^2}} dt\,\frac{\lambda}{1+t\lambda}\,\sum_{k=1}^\infty\theta\!\left(\frac{1}{j_{1,k}^2}-t\right). \label{eq:logW-Stieltjes}
\end{align}
This shows that $\log W(\lambda)$ is a Stieltjes function with $R=j_{1,1}^2$. The spectral measure is again given by a sum of step functions. 

Applying the same argument to the finite $N$ result of the circular Wilson loop \eqref{eq:W-circular}, one can show that its logarithm is also a Stieltjes function.

Only the logarithm of the circular Wilson loop is a Stieltjes function. At strong coupling, the circular Wilson loop $W(\lambda) \sim \exp ( \sqrt{\lambda})$, which is not compatible with a subtracted dispersion relation.
\subsubsection{Bremsstrahlung}
The Bremsstrahlung function is related to the logarithm of the circular Wilson loop through a logarithmic derivative. By differentiating the Stieltjes representation \eqref{eq:logW-Stieltjes}, we obtain
\begin{equation}
B(\lambda)
=\frac{1}{2\pi^2} \lambda \partial_\lambda\log W
=\frac{1}{2\pi^2}\int_0^\infty dt\,\frac{\lambda}{1+t\lambda}\,
\sum_{k=1}^\infty \delta\!\left(1-t \,j_{1,k}^2\right).
\label{eq:B-Stieltjes}
\end{equation}
In this case, the spectral measure is distributional but still positive. Again, this property also holds at finite $N$.
\subsubsection{Flux tube symbol at $\ell=0$} 
There exist closed-form expressions for the Fredholm determinants of the flux tube symbol for $\ell =0,1,2$. Using the closed-form expression \eqref{eq:ftD0} for $\ell = 0$, we obtain the dispersive representation
\begin{equation}
    \log D_0^{\mathrm{f.t.}}(\lambda) = \int_0^{\tfrac{1}{\pi^2}}dt \, \frac{\lambda}{1+t\lambda} \sum_{k=1}^\infty\left[\frac{3}{8}\,\theta\!\left(\frac{1}{4\pi^2(k-\tfrac12)^2}-t\right)
    -\frac18\,\theta\!\left(\frac{1}{4\pi^2k^2}-t\right)\right].
\end{equation}
The measure is positive for all $t \geq 0$ since $(k-\tfrac{1}{2})^2 < k^2$,
the first step function is active whenever the second is, and the residual
coefficient $\tfrac{3}{8} - \tfrac{1}{8} = \tfrac{1}{4}$ is positive.

There also exist closed form expressions for $\log D^{\mathrm{f.t.}}_1$ and $\log D^{\mathrm{f.t.}}_2$. It can be checked that these admit a once-subtracted dispersion representation, but the measure is no longer positive. Indeed, it is possible to show that any $\log D_\ell$ coming from an exponentially decaying symbol function is not compatible with Stieltjes positivity for $\ell \geq 1$. One can show this using Krein's theorem mentioned in Section~\ref{sec:stieltjes}.
\subsubsection{BMN operators}
By \eqref{eq:BMNdimensions}, each BMN scaling dimension is a linear combination of the $\sqrt{1+n^2\lambda'}$ with non-negative occupation numbers $N_n$ as coefficients. Since each $\sqrt{1+n^2\lambda'}$ is a Stieltjes function of $\lambda'$ and the Stieltjes property is preserved under positive linear combinations, every BMN scaling dimension is itself Stieltjes in the effective coupling $\lambda'$.
\subsection{Numerical Stieltjes tests}
\label{sec:numerical-tests}
No closed-form expressions are available for the other observables listed in Section~\ref{sec:observables}. Instead of checking positivity directly, we
test the equivalent Hankel matrix conditions \eqref{eq:hankel-constraints} on the perturbative coefficients. For each observable we proceed as follows:
\begin{enumerate}[label=(\roman*)]
\item We compute $a_n$ to high loop orders from the integrability-based representations.
\item The perturbative coefficients are computed with
high-precision numerics. Unless otherwise stated, we work with 200-digit precision.
\item We verify the moment problem constraints \eqref{eq:hankel-constraints}. The radius of convergence $R$ is known analytically for all observables. 
\end{enumerate}
The exception is the Konishi anomalous dimension, which is not known exactly. We will comment on this case in Subsection~\ref{sec:Konishi}.

It is an interesting question how stringent these moment problem constraints are. We illustrate this in Section~\ref{sec:applications}, where we show to what accuracy the first 99 coefficients of $\Gcusp$ determine the 100th. We list our results for each observable in turn.
\subsubsection{Cusp anomalous dimension} \label{sec:cuspnumeric}
For the cusp anomalous dimension $\Gcusp(\lambda)$ we computed $a_n$ numerically to 100 loops. The radius of convergence is known to be $R=\pi^2$. The Hankel matrix constraints in \eqref{eq:hankel-constraints} are all satisfied at $N=100$. In Section~\ref{sec:applications} we show that the Hankel matrix constraints fix the first 56 significant digits of the 100th perturbative coefficient. This is strong evidence that $\Gcusp$ is Stieltjes.
\subsubsection{Tilted cusp anomalous dimension}
The tilted cusp anomalous dimension $\Gamma(\alpha)$ yields the cusp anomalous dimension $\Gcusp$ at $\alpha=\pi/4$, the octagon anomalous dimension $\Goct$ at $\alpha=0$, and the hexagon anomalous dimension at $\alpha=\pi/3$.    

Since we have already investigated the first two in Section~\ref{sec:octagonanomaloussec} and Section~\ref{sec:cuspnumeric}, we are left to check the Stieltjes property for the hexagon anomalous dimension. For this, we computed the perturbative coefficients of $\Gamma(\pi/3)$ to 100 loops, and verified the Hankel constraints \eqref{eq:hankel-constraints} to this order.

For generic $\alpha$, the perturbative coefficients are computationally more expensive. We verified the constraints
\eqref{eq:hankel-constraints} for the first 20 loops\footnote{We thank Lance Dixon for providing us with this data.} on a uniform discretisation of
$\alpha\in[0,\pi/2]$ with 100 points. The moment problem constraints are satisfied
everywhere. 

\subsection{Observables for which the Stieltjes property does not hold}
\label{sec:failures}
We now turn to observables that do not satisfy the Stieltjes property.
\subsubsection{Konishi anomalous dimension}\label{sec:Konishi}
The full Konishi anomalous dimension is given up to ten loops in \cite{Marboe_2015}. Using this result, we checked that the moment problem constraints \eqref{eq:hankel-constraints} are violated starting at 7 loops with the $H_1^7$ constraint failing first. It is interesting to ask whether the Stieltjes property would survive if we drop the wrapping corrections. Indeed, just using the contribution from the asymptotic Bethe ansatz, $H_1^7$ is positive semidefinite. However, even without the wrapping corrections, the Konishi anomalous dimension fails the moment-problem constraint starting at 14 loops through $H_1^{14}$.

\subsubsection{Octagon correlation function}\label{sec:failuresoctagon} 
The octagon correlation function introduced in Section~\ref{sec:fredholm} depends on two kinematical variables $y$ and $\xi$. We focus on the octagon with zero bridge length and ask whether $-\log D_0(y,\xi)$ is Stieltjes. The determinant $D_0$ itself is exponentially suppressed at strong coupling and so cannot be Stieltjes. We therefore study $-\log D_0$, which is positive for $\lambda>0$.

For generic $y$ and $\xi$ it is computationally difficult to expand $-\log D_0(y,\xi)$ to high orders in perturbation theory. For the octagon symbol function the expansion \eqref{eq:logD0-weak} can be expressed through ladder diagrams \cite{Belitsky_2020}. Using this formalism we compute twelve perturbative orders and test the $N=12$ Hankel constraints \eqref{eq:hankel-constraints} for Euclidean and Lorentzian kinematics separately.

For Euclidean kinematics, we take $iy \in [0,\pi]$ and $\xi \geq 0$. We find that the Hankel constraints fail at generic points, holding only for small $\xi$ at generic $y$. It remains unclear whether these surviving points are genuine or merely an artifact of truncating the constraints at twelve loops.

For real $y$ and $\xi$ the configuration is Lorentzian. Here we again find that the Hankel constraints fail at generic points, holding only for small $y$ and $\xi$ and asymptotically for $y\gg\xi$. In the latter region $-\log D_0(y,\xi)$ is controlled by the octagon anomalous dimension \eqref{eq:Goct-closed}, which we proved to be Stieltjes in Section~\ref{sec:analytic-proofs}. In both regions it is unclear what happens once more perturbative coefficients are included in the Hankel constraints. The octagon might be Stieltjes only in the strict limit $y\to\infty$ at fixed $\xi$, and the region for small $y,\xi$ might disappear as further constraints are imposed.

A more decisive test is possible at the symmetric point $y=\xi=0$ corresponding to $z=\bar{z}=-1$, where the octagon's equations simplify and we can compute 100 perturbative coefficients. There the Hankel constraints are satisfied.

In summary, the Stieltjes property does not hold at generic kinematical points, and it is an open question whether any region exists where it does. We note in addition that real $y$ and $\xi$ cover only part of the Lorentzian kinematics. It is intriguing that the symmetric point satisfies the Hankel constraints to 100 loops. By contrast, the weaker positivity property of complete monotonicity holds at every point we tested, as we show in Appendix~\ref{app:CM}.

\section{Applications: Pad\'e bounds and the moment bootstrap}
\label{sec:applications}

In this section, we develop two applications of the Stieltjes property for observables. We illustrate both applications on the cusp anomalous dimension.
\subsection{Non-perturbative bounds from Pad\'e approximants}
As mentioned in Section~\ref{sec:stieltjes}, Pad\'e approximants of Stieltjes functions have very strong properties. In particular, the Pad\'e approximants $P_N^{N+1}$, $P_N^N$ bound the observable $\mathcal{O}(\lambda)$:
\begin{equation} \label{eq:Padebounds}
    P_N^{N+1}(\lambda) \geq \mathcal{O}(\lambda) \geq P_N^N(\lambda), \qquad \lambda \geq 0.
\end{equation}
This is particularly interesting, because the Pad\'e approximants $P_M^N$ are constructed only from the perturbative expansion of $\mathcal{O}(\lambda)$. Hence, Stieltjes positivity implies that perturbative data can be used to obtain non-perturbative bounds on observables. 

A common practice is to combine Pad\'e approximants with a conformal remapping \cite{Bern_2007,Costin:2021bay}. In this way, the $\sqrt{\lambda}$ behaviour at strong coupling can be correctly captured by the Pad\'e approximants. However, one loses the rigorous bounds of the Pad\'e approximants \eqref{eq:Padebounds}. We are not aware of any literature discussing the connection between Stieltjes functions and conformal mappings. 
\subsection{Moment problem bootstrap}\label{sec:momentbootstrap}
The perturbative expansion coefficients of Stieltjes functions are related to moments of a positive measure. As mentioned in Subsection~\ref{sec:moment-problems}, this implies a set of Hankel matrix constraints given in \eqref{eq:hankel-constraints}. In Section~\ref{sec:positivity}, we used these to numerically test if a given observable is Stieltjes. If we turn this around and assume that an observable is Stieltjes, one can use the Hankel matrix constraints to bound the space of possible perturbative coefficients. Knowing the first $a_1, a_2, \dots, a_N$ expansion coefficients puts numerical bounds on all higher coefficients $a_{N+1}, a_{N+2}, \dots$. 
\subsection{Example: Cusp anomalous dimension}
We illustrate the two applications on the cusp anomalous dimension $\Gcusp(\lambda)$. We did not prove the Stieltjes property of $\Gcusp$, but we will assume that it holds throughout this section. Of course, the same techniques can be applied to any of the Stieltjes observables in Section~\ref{sec:positivity}.

\begin{figure}[t]
\centering
\begin{subfigure}{0.48\linewidth}
\centering
\includegraphics[width=\linewidth]{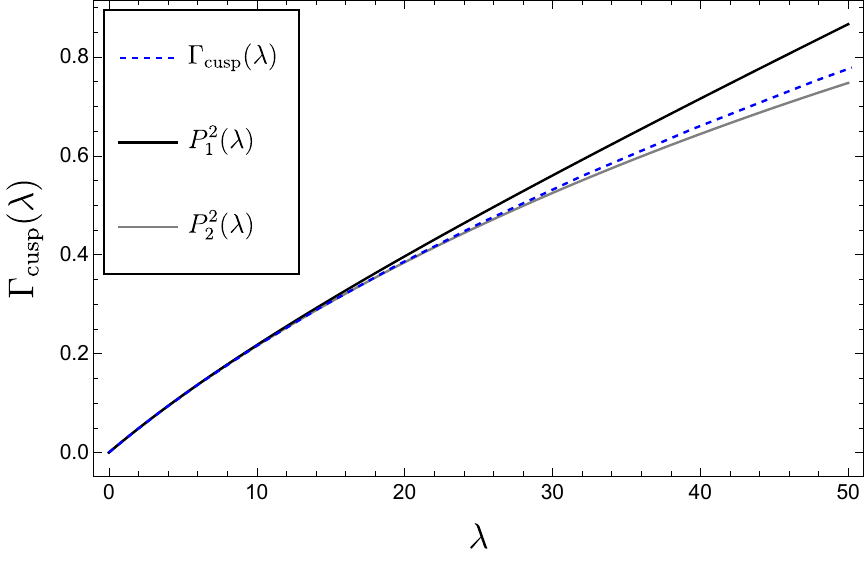}
\caption{}
\label{fig:pade-weak}
\end{subfigure}
\hfill
\begin{subfigure}{0.48\linewidth}
\centering
\includegraphics[width=\linewidth]{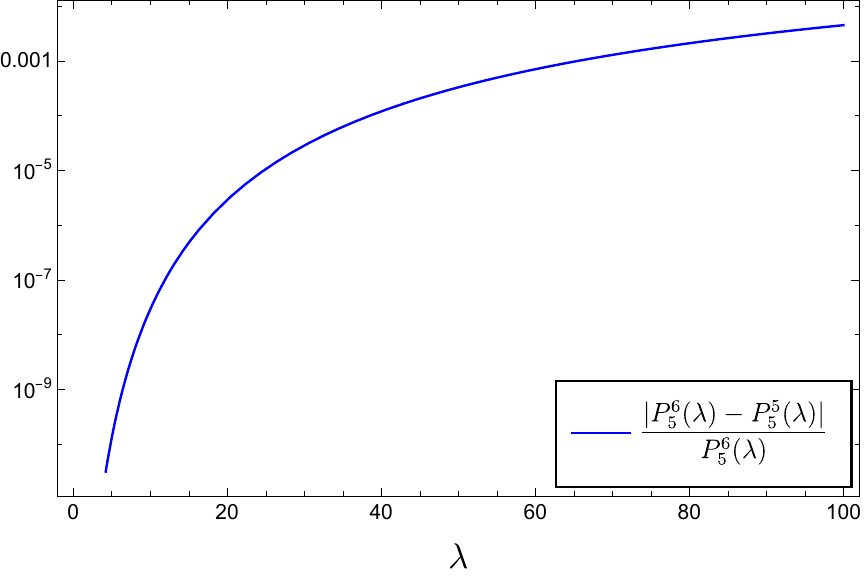}
\caption{}
\label{fig:pade-strong}
\end{subfigure}
\caption{(a): We compare the $P_2^2$ and $P_1^2$ approximants, obtained from only four perturbative coefficients, and the exact value of $\Gcusp$. (b): We show the normalized bound on the error resulting from the $P_5^6$ and $P_5^5$ Pad\'e approximants, obtained from only eleven perturbative coefficients, in a logarithmic plot.
} 
\label{fig:cusp-pade}
\end{figure}

The cusp anomalous dimension has been computed in perturbation theory up to four loops \cite{Bern_2007}. Using these data, we can construct the $P_2^2$ and $P^2_1$ Pad\'e approximants. Pad\'e approximants of this series were already discussed in \cite{Bern_2007}. However, here we want to emphasize that the Stieltjes property allows for rigorous statements. The comparison of the Pad\'e approximants and the exact result for $\Gcusp$ is shown in Figure~\ref{fig:pade-weak}. The radius of convergence of the perturbative series is $\pi^2$. Hence, Pad\'e works well in extrapolating the perturbative information. The exact curve for $\Gcusp(\lambda)$ was obtained by truncating the matrix \eqref{eq:Kij} to a $10\times 10$ block, evaluating its elements numerically, computing $\Gcusp$ for twelve values of $\lambda$ between 0 and 50, and interpolating the result. 

In order to see how the convergence of the Pad\'e approximants improves, we also construct the $P_5^6$ and $P_5^5$ Pad\'e approximants using the first eleven perturbative coefficients. We show the relative difference between the Pad\'e approximants in a logarithmic plot in Figure~\ref{fig:pade-strong}. For $\lambda \in (0,100)$, the relative bound on the error is given by
\begin{equation}
    \frac{|P_5^6(\lambda)-P_5^5(\lambda)|}{P_5^6(\lambda)} \leq 0.005 \qquad \forall \lambda \in (0,100).
\end{equation} 
Now, we apply the moment bootstrap from Subsection~\ref{sec:momentbootstrap} to the cusp anomalous dimension. The radius of convergence for $\Gcusp(\lambda)$ is $R= \pi^2$. Writing $\Gcusp(\lambda) = \sum_{n=1}^\infty (-1)^{n+1}a_n \lambda^n$, the four loop result \cite{Bern_2007} yields the following bounds for $a_5$ and $a_6$:
\begin{equation}\label{eq:a5a6region}
1.002\times 10^{-7}<a_5<1.215\times 10^{-7},
\qquad 7.989\times 10^{-9}<a_6<9.929\times 10^{-9},
\end{equation}
to be compared with the actual values
$a_5=1.079\times 10^{-7}$ and $a_6= 8.314\times 10^{-9}$. The bounds tighten as more expansion coefficients are included. To illustrate this we repeat the above procedure with eleven perturbative coefficients as input. The bounds on $a_{12},a_{13}, a_{14}$, and $a_{15}$ are:
\begin{align}
   &3.21008 \times 10^{-15} <a_{12}<3.21021 \times 10^{-15}\\
   &2.88697\times 10^{-16} <a_{13}< 2.8877\times 10^{-16}\\
    &2.61879 \times 10^{-17} <a_{14}<2.62108 \times 10^{-17}\\
    &2.39276 \times 10^{-18}<a_{15}< 2.39816 \times 10^{-18}.
\end{align}
Using this formalism, we can also comment on how stringent the numerical verification of the Stieltjes property in Subsection~\ref{sec:numerical-tests} is. For the cusp anomalous dimension, we computed 100 perturbative coefficients and verified the Hankel matrix constraints \eqref{eq:hankel-constraints}. We want to investigate to what accuracy the first 99 perturbative coefficients determine the 100th coefficient $a_{100}$. Let $a_{100}^{\text{upper/lower bound}}$ be the upper and lower bound on $a_{100}$ that is compatible with positive semidefiniteness of the Hankel matrices \eqref{eq:hankel-constraints}. We find the following normalized bound
\begin{equation}
    \frac{a_{100}^{\text{upper bound}}-a_{100}^{\text{lower bound}}}{a_{100}^{\text{upper bound}}} \approx 1.05 \times 10^{-56}. 
\end{equation}
Hence, the Stieltjes property determines the first 56 significant digits of $a_{100}$. The fact that the cusp anomalous dimension lies exactly within the window allowed by positivity is a non-trivial test of the Stieltjes property.

We can compare this bootstrap analysis with the previously constructed Pad\'e approximants. For example, another way of getting numerical estimates for the perturbative coefficients $a_5$ and $a_6$ by using $a_1, a_2, a_3, a_4$ is to use the Pad\'e approximant $P_2^2(\lambda)$. Expanding this function at $\lambda=0$ yields estimates $a_5^P, a_6^P$ for $a_5$ and $a_6$. This yields
\begin{equation}
    a_5^P = 9.994 \times 10^{-8} \qquad a_6^P = 6.615 \times 10^{-9}.
\end{equation}
They are numerically close to the actual values of $a_5$ and $a_6$. However, importantly they do not lie in the allowed region of \eqref{eq:a5a6region}. In this sense, the bootstrap is a better way to estimate higher unknown perturbative coefficients than the Pad\'e approximants. 

\section{Strong-coupling expansion from the dispersion relation}
\label{sec:strong}

In this section, we discuss how to obtain the strong-coupling expansion from the dispersion relation \eqref{eq:once-subtracted}. For this discussion, we do not need to assume positivity of the spectral density. The results will generally hold for a once-subtracted dispersion relation.
\subsection{Mellin--Barnes representation}
The weak-coupling expansion of \eqref{eq:once-subtracted} is trivially
obtained by expanding the integrand in powers of $t\lambda$. The
strong-coupling expansion is more subtle. The naive expansion
$\lambda/(1+t\lambda)=\sum_{n=0}^\infty(-1)^n(t\lambda)^{-n}/t$ leads to problems when exchanging integration and summation and cannot reproduce the $\sqrt\lambda$ growth
seen in the observables listed in Section~\ref{sec:observables}. However, this can be solved by using a Mellin--Barnes
representation\footnote{We thank Gregory Korchemsky for suggesting this.}. We write,
\begin{equation}
\frac{\lambda}{1+t\lambda}
=\int_{\delta-i\infty}^{\delta+i\infty}\frac{dj}{2\pi i}
\frac{\pi}{\sin\pi j}\,\lambda^j t^{j-1},
\qquad 0<\mathrm{Re}\,\delta<1.
\label{eq:mb-identity}
\end{equation}
Using this identity, \eqref{eq:once-subtracted} becomes
\begin{equation}
\mathcal{O}(\lambda)
=\int_{\delta-i\infty}^{\delta+i\infty}\frac{dj}{2\pi i}
\frac{\pi}{\sin\pi j}\,\lambda^j\,\hat\nu(j),
\qquad
\hat\nu(j)=\int_0^{1/R} dt\,t^{j-1}\nu(t).
\label{eq:mb-master}
\end{equation}
The spectral measure $\nu(t)$ enters the integrand through its Mellin transform $\hat{\nu}(j)$. For $|\lambda|<R$, the Mellin--Barnes contour can be closed to the right. Here, the Mellin transform does not have any poles, and the integral picks up the residues at $j=1,2,3,\dots$ from the $\pi/\sin( \pi j)$ contribution. This reproduces the convergent weak-coupling expansion.  

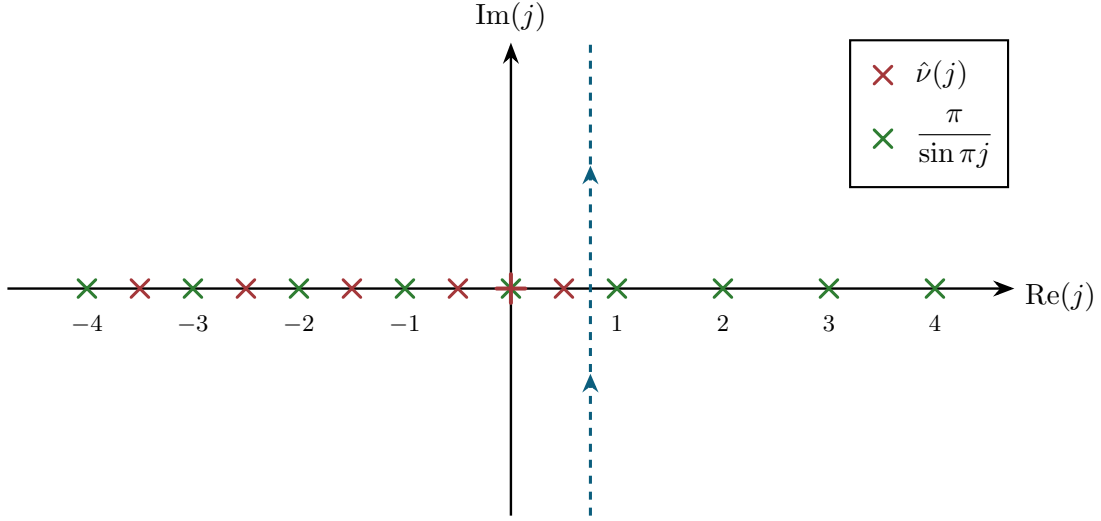
\begin{figure}[t]
\centering
\begin{tikzpicture}[x=1.4cm,y=1.0cm]

  \draw[-{Stealth[length=8pt,width=6pt]}, line width=0.9pt]
        (-4.75,0) -- (4.75,0) node[right, anchor=west, yshift=-3pt]{$\mathrm{Re}(j)$};
  \draw[-{Stealth[length=8pt,width=6pt]}, line width=0.9pt]
        (0,-3.0) -- (0,3.25) node[above]{$\mathrm{Im}(j)$};

  \foreach \n in {-4,-3,-2,-1,0,1,2,3,4}{\cross{forest}{\n,0}}

  \foreach \n in {0.5,-0.5,-1.5,-2.5,-3.5}{\cross{brickred}{\n,0}}

  \foreach \n in {-4,-3,-2,-1,1,2,3,4}
     {\node[font=\footnotesize, anchor=north] at (\n,-0.22) {$\n$};}

  \draw[tealblue, dashed, line width=1.15pt,
        postaction={decorate, decoration={markings,
          mark=at position 0.30 with {\arrow[tealblue]{Stealth[length=7pt,width=6pt]}},
          mark=at position 0.74 with {\arrow[tealblue]{Stealth[length=7pt,width=6pt]}}}}]
        (0.75,-3.0) -- (0.75,3.25);

  \draw[brickred, line width=1.5pt, line cap=round]
     ([shift={(-5.4pt,0)}]0,0) -- ([shift={(5.4pt,0)}]0,0)
     ([shift={(0,-5.4pt)}]0,0) -- ([shift={(0,5.4pt)}]0,0);

  \draw[black, line width=0.8pt] (3.20,1.34) rectangle (4.69,3.28);
  \cross{brickred}{3.51,2.82}
  \node[anchor=west] at (3.72,2.82) {$\hat\nu(j)$};
  \cross{forest}{3.51,1.98}
  \node[anchor=west] at (3.72,1.98) {$\dfrac{\pi}{\sin\pi j}$};

\end{tikzpicture}
\caption{Analytic structure of the Mellin--Barnes integrand
\eqref{eq:mb-master}. Green crosses mark the simple poles of
$\pi/\sin\pi j$ at integers $j\in\mathbb{Z}$, red crosses mark the
poles of $\hat\nu(j)$.
Closing the contour to the right produces the convergent
weak-coupling expansion, closing to the left produces the asymptotic strong-coupling expansion.}
\label{fig:mellin}
\end{figure}

For $|\lambda|>R$ the contour can no longer be closed to the right. Closing the integration contour to the left requires an analytic continuation of $\hat{\nu}(j)$ to $\operatorname{Re}j<1$. There will be poles at $j=0,-1,-2, \dots$ coming from $\pi/\sin(\pi j)$ and poles coming from $\hat{\nu}(j)$. Since $\hat{\nu}(j)$ yields the perturbative coefficients $a_j$ for $j=1,2,3,\dots$, one can think of $\hat{\nu}(j)$ as the analytic continuation of the perturbative coefficients. In general, the integrand might not vanish for $\operatorname{Re}j \to - \infty$. This will be related to an asymptotic strong-coupling expansion and non-perturbative contributions. The analytic structure of the integrand is summarized in Figure~\ref{fig:mellin}.
We illustrate this procedure in the following example.
\subsection{Example: Octagon anomalous dimension}
\label{sec:mb-octagon}
We illustrate how to obtain the strong-coupling expansion of the octagon anomalous dimension from the dispersive representation \eqref{eq:Goct-Stieltjes}. First, we compute the Mellin transform of the spectral measure
\begin{equation}
    \hat{\nu}(j) = \int_0^{\tfrac{1}{\pi^2}} dt\, t^{j-1} \sum_{n=0}^\infty \theta\left( \frac{1}{4 \pi^2(n+1/2)^2}-t\right) = \frac{(1-2^{-2j})\pi^{-2j}}{j} \zeta(2j).
\end{equation}
The Mellin transform has a simple pole at $j=1/2$ and a spurious pole at $j=0$. Plugging this into the Mellin--Barnes representation yields
\begin{equation} \label{eq:Goctmellin}
    \Goct(\lambda) = \int_{\delta-i \infty}^{\delta+i \infty} \frac{dj}{2\pi i} \, \frac{\pi}{\sin \pi j} \lambda^j \frac{(1-2^{-2j})\pi^{-2j}}{j} \zeta(2j) \qquad \frac{1}{2}<\operatorname{Re}\delta < 1.
\end{equation}
In order to study the strong-coupling expansion of this integral, we are interested in the behavior of the integrand for $\operatorname{Re}j<1$. For this it is convenient to rewrite $\zeta(2j)$ using its reflection identity. This yields 
\begin{equation}
      \Goct(\lambda) = \int_{\delta-i \infty}^{\delta+i \infty} \frac{dj}{2\pi i} \, \lambda^j \frac{(2^{2j}-1)}{j} \Gamma(1-2j)\zeta(1-2j) \qquad \frac{1}{2}<\operatorname{Re}\delta < 1.
\end{equation}
For $\operatorname{Re}j<1$, the integrand only has simple poles at $j=1/2$ and $j=0$. All poles of $\pi/\sin(\pi j)$ in \eqref{eq:Goctmellin} for negative integers are canceled by zeros of $\hat{\nu}(j)$. Shifting the integration contour to the left yields
\begin{equation} \label{eq:OctagonMellinPert}
    \Goct(\lambda) = \frac{1}{2}\sqrt{\lambda} - \log 2 + \int_{\delta-i\infty}^{\delta + i \infty} \frac{dj}{2\pi i} \,\frac{1}{j} \lambda^j (2^{2j}-1) \Gamma(1-2j) \zeta(1-2j) \qquad \operatorname{Re}\delta <0.
\end{equation}
The remaining contour integral does not vanish, and does not contribute to the perturbative part of the strong-coupling expansion because the integrand has no poles. This is related to non-perturbative contributions to the strong-coupling expansion. To see this, we shift the integral to $\operatorname{Re} \delta \to - \infty$ and evaluate it using a saddle point approximation. It is convenient to change variables from $j$ to $-j$. The integrand reads
\begin{equation}
    g(j) = \frac{1}{2\pi i} \frac{1}{j} \lambda^{-j} (1-2^{-2j}) \Gamma(1+2j) \zeta(1+2j).
\end{equation}
For large $j$, we can rewrite the $\zeta(1+2j)$ as an infinite sum and expand the $\Gamma(1+2j)$ using Stirling's approximation. Keeping only the first term in this expansion yields
\begin{equation}
    g(j) \sim \frac{1}{2\pi i} \frac{1}{j}  \sum_{n=1}^\infty \frac{(-1)^{n+1}}{n}\exp \Big[ j(2 \log j+\log 4 -2 -2\log n - \log \lambda) \Big] \left(2\sqrt{\pi} \sqrt{j}+ \mathcal{O}\big(j^{-1/2} \big)\right).
\end{equation} 
Each summand has a saddle point located at $j= n \tfrac{\sqrt{\lambda}}{2}$. Evaluating the integral using the saddle point approximation yields
\begin{equation}
    \Goct(\lambda) = \frac{1}{2}\sqrt{\lambda} - \log 2 + \sum_{n=1}^\infty \frac{(-1)^{n+1}}{n} e^{-n \sqrt{\lambda}} \left( 1+ \mathcal{O}(\lambda^{-1/2})\right).
\end{equation}
In principle, subleading terms can be obtained by including higher terms in Stirling's approximation and the saddle point approximation. However, it turns out that all of these vanish for the octagon anomalous dimension. To see this, note that
\begin{equation}
    \Goct(\lambda) = \log \cosh \frac{\sqrt{\lambda}}{2} = \frac{1}{2}\sqrt{\lambda}-\log 2 + \sum_{n=1}^\infty \frac{(-1)^{n+1}}{n} e^{-n\sqrt{\lambda}}.
\end{equation}
The analogous computation for the circular Wilson loop is
significantly more intricate.
We present it in Appendix~\ref{app:circular}.
\subsection{From weak to strong coupling}
The Mellin transform $\hat{\nu}(j)$ can be seen as an analytic continuation of the perturbative coefficients $a_n$. This raises the question of whether the strong-coupling expansion can be extracted from only considering weak-coupling information. We will briefly comment on this below. 

A way to approximate the analytic continuation of $\hat{\nu}(j)$ is to take the known values of $\hat{\nu}(n) = a_n$ for $n=1, \dots, N$ and construct a multi-point Pad\'e approximant. Here $a_n$ are the perturbative coefficients. The poles of the resulting Pad\'e approximant encode information about the strong-coupling expansion of the observable.

\begin{figure}[t]
\centering
\begin{subfigure}{0.48\linewidth}
\centering
\includegraphics[width=\linewidth]{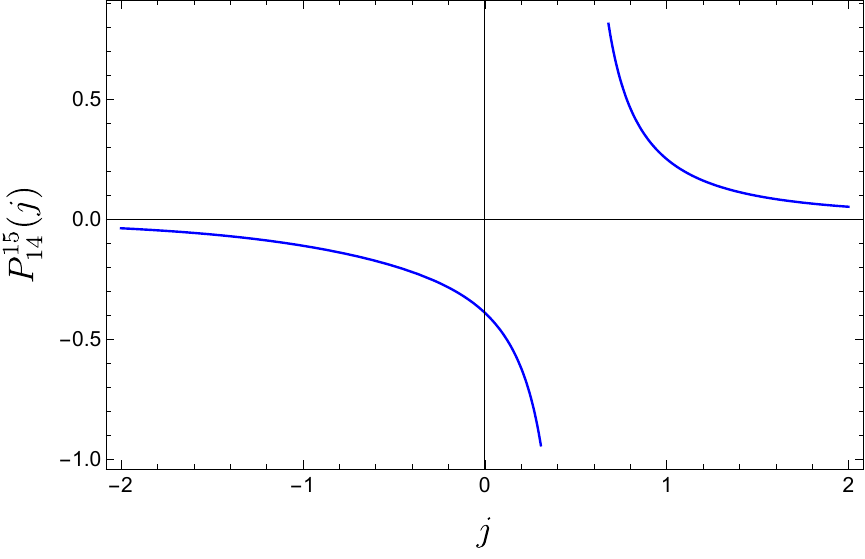}
\caption{}
\label{fig:multi30}
\end{subfigure}
\hfill
\begin{subfigure}{0.48\linewidth}
\centering
\includegraphics[width=\linewidth]{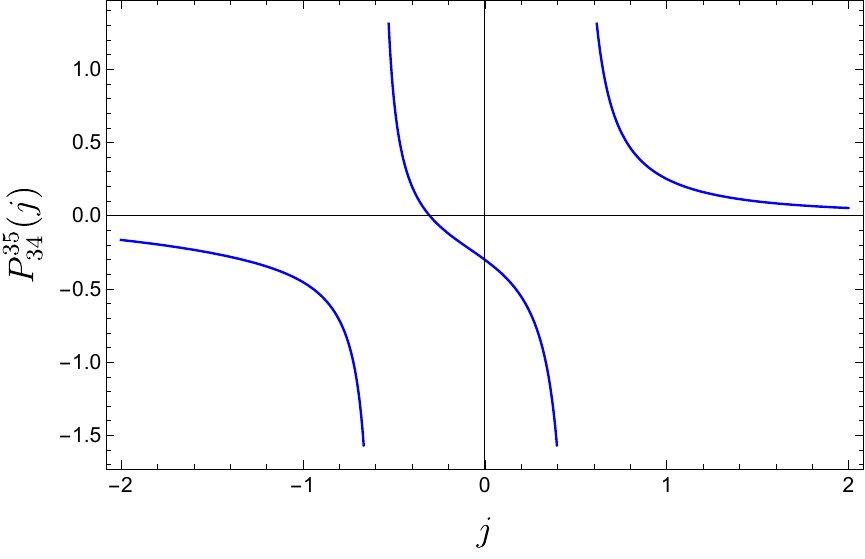}
\caption{}
\label{fig:multi70}
\end{subfigure}
\caption{(a) The $P_{15}^{14}$ multi-point Pad\'e approximant for $\hat{\nu}(j)$. The approximant correctly captures a pole close to $j=1/2$.  (b) The $P_{35}^{34}$ multi-point Pad\'e approximant for $\hat{\nu}(j)$. The approximant correctly captures a pole close to $j=1/2$ and $j=-1/2$.
} 
\label{fig:multipointpade}
\end{figure}

We will illustrate this on the cusp anomalous dimension. For this, we introduce the quantity $\gamma(\lambda)=\Gcusp(\pi^2 \lambda)$, which has a unit radius of convergence. This ensures that multi-point Pad\'e approximants constructed from the perturbative coefficients of $\gamma$ are numerically more stable. We construct the $P_{15}^{14}$ and $P_{35}^{34}$ Pad\'e approximants from $N=30$ and $N=70$ perturbative coefficients respectively and show the results for the analytic continuation of $\hat{\nu}(j)$ in Figure~\ref{fig:multipointpade}. Based on the strong-coupling expansion \eqref{eq:Gcusp-strong} there should be simple poles at $j=1/2,\, -1/2, \, -3/2, \, \dots$. The $N=30$ approximant correctly captures a pole in the vicinity of $j=1/2$ but none of the other poles. Increasing the order to $N=70$ captures a pole in the vicinity of $j=-1/2$ as well. We can use the $P_{35}^{34}$ approximant instead of $\hat{\nu}(j)$ in \eqref{eq:mb-master} to approximate the strong-coupling behavior. Translating from $\gamma$ to $\Gcusp$ yields the approximate strong-coupling expansion 
\begin{equation}
    \Gcusp \approx 0.158 \lambda^{0.5009}- 0.301-1.19 \lambda^{-0.590} -\dots.
\end{equation} 
We can compare this to the exact result 
\begin{equation}
    \Gcusp \sim 0.159 \lambda^{0.5}-0.331-0.146 \lambda^{-0.5}-\dots .
\end{equation}
The method of analytically continuing $a_n$ using multi-point Pad\'e approximants works qualitatively. Since the strong-coupling expansion is only asymptotic, we expect the Mellin transform $\hat{\nu}(j)$ to diverge factorially as $j \to - \infty$. This behaviour can never be captured by Pad\'e approximants and constitutes a fundamental limitation of the use of Pad\'e approximants for the analytic continuation. 
\subsection{Discussion}
The strong-coupling expansion of \eqref{eq:mb-master} is controlled by the poles of the integrand for $\operatorname{Re}j<1$. These come from $\pi/\sin (\pi j)$ and $\hat{\nu}(j)$. In particular, the poles of the Mellin transform $\hat{\nu}(j)$ are controlled by the behavior of the spectral density $\nu(t)$ for $t\to 0$. For the octagon anomalous dimension, one has 
\begin{equation}
    \nu(t) \sim \frac{1}{2\pi} \frac{1}{\sqrt t}, \qquad t \to 0.
\end{equation}  
This leads to the pole at $j=1/2$, which in turn is responsible for the $\sqrt{\lambda}$ behavior at strong coupling. 

We expect the following structure to hold at strong-coupling for the observables listed in Section~\ref{sec:observables}:
\begin{equation}
    \mathcal{O}(\lambda) \sim A_0 \sqrt{\lambda} + A_1 \log \lambda + A_2 + \sum_{k=1}^\infty \frac{A_{k+2}}{\sqrt{\lambda^k}}, \qquad \lambda \to \infty
\end{equation}
If this expansion holds, the spectral measure must have the following behavior
\begin{equation}
    \nu(t) \sim B_0 \frac{1}{\sqrt{t}}+ B_1 + \sum_{n=1}^\infty B_{n+1} t^{(2n-1)/2}, \qquad t \to 0.
\end{equation}
The constant piece $B_1$ will lead to a simple pole at $j=0$ in the Mellin transform $\hat{\nu}(j)$,
which will combine with the simple pole from the $\pi/\sin (\pi j)$ to give a double pole that reproduces the $\log \lambda$ term. The other terms give simple poles at $j=1/2, -1/2, -3/2, \dots $. 

For the observables listed in Section~\ref{sec:observables}, the strong-coupling expansion receives non-perturbative contributions. As shown in Subsection~\ref{sec:mb-octagon} for the octagon anomalous dimension, these are due to the Mellin--Barnes integration not vanishing for $\operatorname{Re}j \to - \infty$. The non-perturbative contributions can be computed from a saddle point analysis. 

The non-perturbative contributions can also be seen in the spectral measure $\nu(t)$. For the octagon anomalous dimension, it can be checked that 
\begin{equation}
    \nu_{\mathrm{oct}}(t) = \frac{1}{2\pi \sqrt{t}} + \frac{1}{\pi} \sum_{n=1}^\infty \frac{(-1)^n}{n} \sin\Big( \frac{n}{\sqrt{t}} \Big).
\end{equation}
Every non-perturbative contribution leaves its imprint as a non-analytic, highly oscillatory piece at the origin. We expect this structure to hold generally for the observables considered.

\section{Conclusion}
\label{sec:conclusion}

In this paper we showed that several exactly known observables in planar
$\mathcal{N}=4$ sYM are analytic functions of the 't~Hooft coupling and admit a
once-subtracted dispersion representation over a positive measure. Our findings
are summarized in Table~\ref{tab:summary}: for each observable we indicate
whether it satisfies the Stieltjes property, and, in those cases where we could
establish this only perturbatively, the loop order to which we verified it. The Stieltjes property for the octagon correlation function is not yet satisfactorily understood. We argued that for generic kinematical points it is not a Stieltjes function. However, it is not clear if there is a kinematical region for which the Stieltjes property holds. 
\begin{table}[t] 
\centering
\begin{tabular}{lcl}
\hline
Observable  & Stieltjes & Verified \\[4pt]
\hline
Octagon anomalous dimension  & \checked & analytically\\[4pt]
Logarithm of circular Wilson loop  & \checked & analytically \\[4pt]
Bremsstrahlung  & \checked & analytically \\[4pt]
BMN scaling dimensions & \checked & analytically \\[4pt]
Cusp anomalous dimension  & \checked & numerically, 100 loops\\[4pt]
Hexagon anomalous dimension  & \checked & numerically, 100 loops \\[4pt]
Tilted cusp anomalous dimension & \checked   & numerically, 20 loops\\[4pt]
Octagon $(y,\xi)$ & $\times$  & --- \\[4pt] 
Konishi  anomalous dimension & $\times$ & --- \\[4pt]
\hline
\end{tabular}
\caption{A summary of our findings regarding the Stieltjes property of observables in planar $\mathcal{N}=4$ sYM.}\label{tab:summary}
\end{table}

The Stieltjes property has two immediate applications. First, Pad\'e
approximants constructed from the perturbative coefficients are guaranteed to
yield rigorous non-perturbative bounds. Second, positivity can be used to
bootstrap the perturbative coefficients themselves. We further showed how the
strong-coupling expansion follows from the dispersive representation, and
commented on how the perturbative weak-coupling coefficients can be used to estimate the strong-coupling
coefficients.

We briefly outline future directions we find interesting to pursue.

\begin{enumerate}
    \item Beyond positivity, the mere fact that these observables are analytic
    functions of the coupling is itself striking. Analyticity in kinematic
    variables is deeply connected to locality~\cite{Eden:1966dnq}. It would be interesting
    to ask whether an analogous structural reason underlies analyticity in the
    coupling. A closely related question is whether the positivity we observe in
    the dispersion relation has an underlying explanation, and what
    characterizes the class of observables for which it holds. At present we
    have no such understanding. 
    A natural starting point is to examine
    the analytic structure of the integrability-based representations such
    as \eqref{eq:Kij} and \eqref{eq:Gcusp}, which only converge for
    $|\operatorname{Im} g| < \tfrac{1}{4}$ and require an analytic continuation. It would be interesting to find a representation valid for all $g \in \mathbb{C}$. 

    \item The strong-coupling expansion of these observables is typically
    asymptotic and not Borel summable. For an observable satisfying the
    Stieltjes property, positivity imposes constraints on the weak-coupling
    perturbative coefficients. It would be interesting to determine whether
    comparable constraints govern the strong-coupling coefficients, and whether
    positivity has anything to say about resurgence analyses of them.

    \item Although the Stieltjes property remains conjectural for several
    observables, applying the bootstrap to concrete examples may already be
    instructive. The recently introduced walking anomalous dimension in \cite{Alday:2026cpk}
    interpolates between the octagon and cusp anomalous dimensions, both of
    which we believe to be Stieltjes functions. Therefore, one might hope that the walking anomalous dimension is Stieltjes and that positivity could be used to
    bootstrap its perturbative coefficients. Currently only a two-loop result is available, so the
    resulting bootstrap bounds on the third loop are not yet very strong.

    \item The strong-coupling expansion can be obtained from the
    Mellin--Barnes representation together with an analytic continuation of the
    weak-coupling coefficients $a_n$. It would be interesting to explore how to
    improve this analytic continuation so as to extract better, or even exact,
    strong-coupling coefficients from perturbative data. 
\end{enumerate}
\subsection*{Acknowledgments}  
I am grateful to Johannes Henn, Gregory Korchemsky, and Prashanth Raman for many discussions during the course of my M.Sc.\ thesis, from which this work is drawn. I also thank Nima Arkani-Hamed, Benjamin Basso, and Valentina Forini for helpful discussions, and Lance Dixon and Johannes Henn for comments on the draft. I thank IPhT Saclay for its hospitality during the completion of part of this work. I acknowledge support from the Deutschlandstipendium and from the Max Planck--IAS--NTU Center for Particle Physics, Cosmology and Geometry. 
\newpage
\appendix
\section{Complete monotonicity in $1/\lambda$}
\label{app:CM}
In Subsection~\ref{sec:analytic-proofs} and Subsection~\ref{sec:numerical-tests}, we argued that several observables listed in Section~\ref{sec:observables} satisfy a Stieltjes positivity property. However, in Subsection~\ref{sec:numerical-tests} we were only able to give numerical evidence, and in Subsection~\ref{sec:failures} we commented on observables that do not satisfy the Stieltjes property. Motivated by \cite{ditsch2026approximatingfeynmanintegralsusing}, we introduce a weaker positivity condition. 
\subsection{Complete monotonicity}
\label{sec:cm-def}
Stieltjes functions are a strict subclass of \emph{completely monotonic
functions}, characterised by the Hausdorff--Bernstein--Widder theorem
\cite{WidderWidder+2015}: a function $f:(0,\infty)\to\mathbb{R}$ is
completely monotonic if and only if it admits a representation
\begin{equation}
f(x)=\int_0^\infty dt\, e^{-xt}\,b(t),\qquad b(t)\ge 0.
\label{eq:BHW}
\end{equation}
An equivalent characterisation is that $f$ is infinitely differentiable and $(-1)^n f^{(n)}(x)\ge 0$ for all $n\ge 0$ and $x>0$. Every Stieltjes function is completely monotonic, which can easily be seen by differentiating the Stieltjes representation under the integral sign. The converse is not true. For example, $f(x)=e^{-x}$ is completely monotonic but not Stieltjes. 
The measure $b$ is the inverse Laplace transform of $f$.

There exists a ``Tauberian'' theorem that relates the asymptotic expansion of the measure $b(t)$ for $t \to \infty$ and the asymptotic expansion of $f(x)$ for $x \to 0$ \cite{feller1}. In particular, for $\rho > 0$ the following are equivalent:
\begin{align} 
    f(x) &\sim x^{-\rho} L\Big(\frac 1 x\Big), \qquad x\to 0 \label{eq:TauberianLaplace}\\
    b(t) &\sim \frac{1}{\Gamma(\rho)} t^{\rho -1} L(t), \qquad t \to \infty. \nonumber
\end{align} 
Here, $L(x)$ is any function such that $\lim_{x \to\infty} L(a x)/L(x) = 1$ for all $a>0$.

The connection between Stieltjes functions and completely monotonic functions has been studied in the context of Feynman integrals in \cite{ditsch2026approximatingfeynmanintegralsusing}.

For observables in planar $\mathcal{N}=4$ sYM, we consider integral representations of the form
\begin{equation} \label{eq:CMobsdef}
    \mathcal{O}(\lambda) = \int_0^\infty dt \, e^{-t/\lambda} b(t), \qquad b(t) \geq 0.
\end{equation}
That is, the observable is completely monotonic in $1/\lambda$. However, by slight abuse of terminology, we will refer to any observable $\mathcal{O}(\lambda)$ obeying \eqref{eq:CMobsdef} as completely monotonic. 
To see that this is a weaker positivity condition than Stieltjes, note that
\begin{equation}
    \int_0^{R^{-1}} dt\,\frac{\lambda}{1+t\lambda}\,\nu(t) = \int_0^{R^{-1}} dt\,\int_0^\infty du\,e^{-u/\lambda}e^{-tu}\nu(t) = \int_0^\infty du\,e^{-u/\lambda}\int_0^{R^{-1}} dt\,e^{-tu}\nu(t).
\end{equation}
Thus every Stieltjes observable also admits \eqref{eq:CMobsdef}. 

Complete monotonicity is interesting for two reasons. Firstly, it is a weaker condition, so the observables discussed in Subsection~\ref{sec:failures} might still satisfy it. Secondly, numerical arguments for positivity can be put on a more rigorous ground, as we will explain later.  

A word of caution: The perturbative expansion of $\mathcal{O}(\lambda)$ is typically sign-alternating. This implies that $\mathcal{O}(\lambda)$ is completely monotonic in $\lambda$ at $\lambda = 0$. However, this property does not necessarily extend to finite $\lambda$. In fact, the observables we are interested in are not completely monotonic in $\lambda$ at finite coupling since they grow as $\sqrt{\lambda}$, which is not completely monotonic in $\lambda$. 
\subsection{A path-integral argument for complete monotonicity}
We show how complete monotonicity in $1/\lambda$ can arise naturally from a path-integral representation of the observable.

First, consider the partition function of $\mathcal{N}=4$ sYM on $S^4$ at finite $N$, which can be calculated using localization \cite{Pestun_2012}:
\begin{equation}
Z(\lambda)=\int d\lambda_1\cdots d\lambda_N\,
e^{-\frac{1}{g^2}\sum_i\lambda_i^2}\,\prod_{i<j}(\lambda_i-\lambda_j)^2.
\label{eq:Zmatrix}
\end{equation}
Differentiating under the integral sign with respect to $x=1/g^2$
brings down negative factors of $\sum_i\lambda_i^2\ge0$, so
$Z$ is completely monotonic in $x$ by inspection. 

The circular Wilson loop admits a similar representation as a matrix integral \cite{Pestun_2012}
\begin{equation}
W(x)
=
\frac{1}{Z}
\int d\lambda_1 \cdots d\lambda_N \,
\frac{1}{N}\sum_{i=1}^Ne^{ \lambda_i} \,
e^{-x \sum_i \lambda_i^2}
\prod_{i<j} (\lambda_i - \lambda_j)^2 .
\label{eq:Wilson}
\end{equation}
To see that this is completely monotonic in $x$, we perform a change of variables $u_i =\lambda_i \sqrt{x}$ both in the numerator and the denominator. This leads to the representation
\begin{equation}
    W(x) = C\int du_1 \cdots du_N \,
\frac{1}{N}\sum_{i=1}^N\,e^{ u_i/\sqrt{x}}
e^{- \sum_i u_i^2}
\prod_{i<j} (u_i - u_j)^2,
\end{equation}
where $C$ is a positive constant. Now, we use the $u_i\to -u_i$ symmetry to rewrite this as
\begin{equation}
W(x)=C\int du_1\cdots du_N\, \frac{1}{N}
\sum_{i=1}^N \cosh(u_i/\sqrt x)\,
e^{-\sum_i u_i^2}\prod_{i<j}(u_i-u_j)^2.
\end{equation}
The $x$-dependence is isolated in $\cosh(a/\sqrt x)
=\sum_{n\ge 0}a^{2n}/(2n)!\,x^{-n}$, which is a positive linear
combination of $x^{-n}$ and hence completely monotonic in $x$. This shows that at finite $N$, the circular Wilson loop is completely monotonic in $1/g^2$. In the planar limit, this leads to complete monotonicity in $1/\lambda$.

In Subsection~\ref{sec:analytic-proofs}, we showed that $\log W$ is Stieltjes, and hence completely monotonic in $1/\lambda$. In fact, this implies complete monotonicity of $W$ in $1/\lambda$ as well, since the exponential of a completely monotonic function is completely monotonic (see also the discussion in \cite{henn2025positivitypropertiesscatteringamplitudes}). 
\subsection{Borel summation as an inverse Laplace transform}
\begin{figure}[t]
\centering
\includegraphics[width=0.55\linewidth]{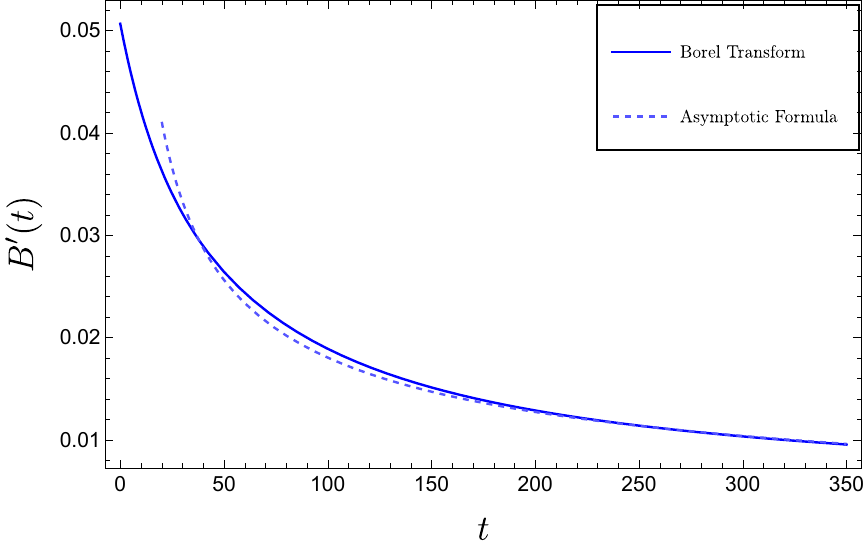}
\caption{$B'(t)$ and its asymptotic expansion for the cusp anomalous dimension. Both agree with each other for $t\gtrsim 150$.}
\label{fig:cuspcm}
\end{figure} 
The representation \eqref{eq:CMobsdef} is directly tied to Borel
summation. Given a perturbative expansion of an observable of the form
$\mathcal{O}(\lambda)=\sum_{n\ge 1}a_n\lambda^n$, its Borel transform is defined by
\begin{equation}
    B(t)=\sum_{n\ge 1}\frac{a_n}{n!}t^n.
\end{equation}
If the perturbative series has a finite radius of convergence, then $B(t)$ is an entire function, and the observable can be recovered by the Laplace transform
\begin{equation}
    \mathcal{O}(\lambda)
=\frac{1}{\lambda}\int_0^\infty dt\,e^{-t/\lambda}\,B(t)
=\int_0^\infty dt\,e^{-t/\lambda}\,B'(t).
\end{equation}
The second form matches \eqref{eq:CMobsdef} with $b(t)=B'(t)$.
Complete monotonicity in $1/\lambda$ is thus equivalent to
\begin{equation}
B'(t)\ge 0,\qquad t>0.
\label{eq:Bprime-positive}
\end{equation}
We can check this condition in perturbation theory by evaluating the Borel transform and truncating the sum at large orders.
At moderate $t$, positivity can be checked directly. However, at large $t$, the truncated series breaks down. This problem can be overcome by using the Tauberian theorem for the Laplace transform \eqref{eq:TauberianLaplace}. Applying this theorem to \eqref{eq:CMobsdef} relates the strong-coupling expansion of the observable to the large $t$ expansion of $B'(t)$. In particular, if
$\mathcal{O}(\lambda)\sim A\sqrt\lambda+B\log\lambda+\cdots$ as
$\lambda\to\infty$, then
\begin{equation} \label{eq:tauberianobs}
B'(t)\sim\frac{A}{\sqrt{\pi t}}+\frac{B}{t}+\cdots,\qquad t\to\infty.
\end{equation}
Our procedure for testing complete monotonicity is as follows. We compute the perturbative expansion for observables following the same procedure as in Subsection~\ref{sec:numerical-tests}. From this, we construct the derivative of the truncated Borel transform $B'(t)$. Furthermore, we use the known strong-coupling expansion of the observable to construct the asymptotic behavior of $B'(t)$ using \eqref{eq:tauberianobs}. Using the perturbative result, we can check for positivity for small values of $t$, and using the Tauberian estimate, we have control over the large $t$ region. 
\subsection{Results}
Since every Stieltjes observable is completely monotonic, all observables shown to be Stieltjes in Section~\ref{sec:analytic-proofs} are completely monotonic as well.
We illustrate complete monotonicity in two examples: the cusp anomalous dimension and the octagon correlation function. For the latter, we discussed in Subsection~\ref{sec:failures} that the Stieltjes property does not hold generically. For the Konishi anomalous dimension, the other observable discussed in Subsection~\ref{sec:failures}, we do not have enough perturbative data available to test for complete monotonicity. 
\subsubsection{Cusp anomalous dimension}
For the cusp anomalous dimension, we computed $B'(t)$ perturbatively for 100 loops. From the strong-coupling expansion \eqref{eq:Gcusp-strong}, we deduce the asymptotic behavior
    \begin{equation}
    B'(t)\sim\frac{1}{2\pi^{3/2}\sqrt t}+\frac{G}{4\pi^{3/2} t^{3/2}}+\cdots, \qquad t\to \infty.
    \end{equation}
The results are shown in Figure~\ref{fig:cuspcm}. Positivity can be seen to hold for small $t$ and in the asymptotic regime for large $t$, thereby establishing positivity for all $t>0$. 
\begin{figure}[t]
\centering
\begin{subfigure}{0.48\linewidth}
\centering
\includegraphics[width=\linewidth]{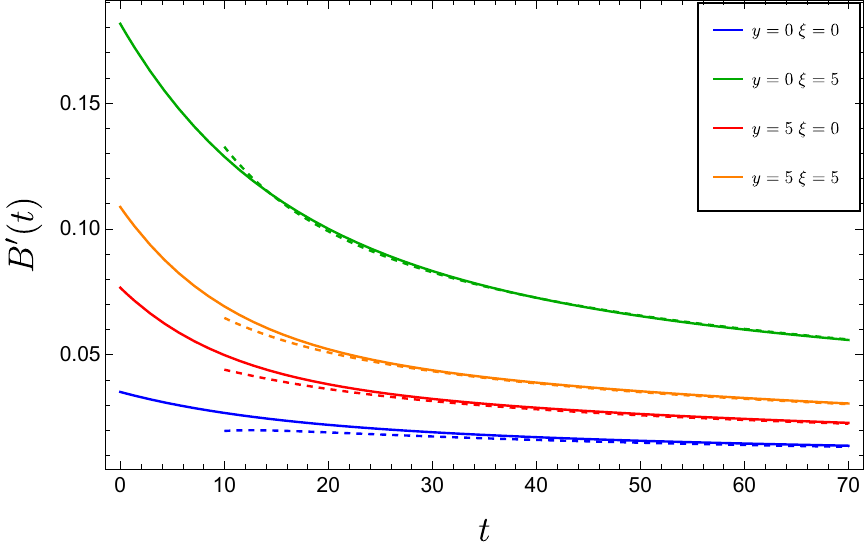}
\caption{}
\label{fig:octCMLorentzian}
\end{subfigure}
\hfill
\begin{subfigure}{0.48\linewidth}
\centering
\includegraphics[width=\linewidth]{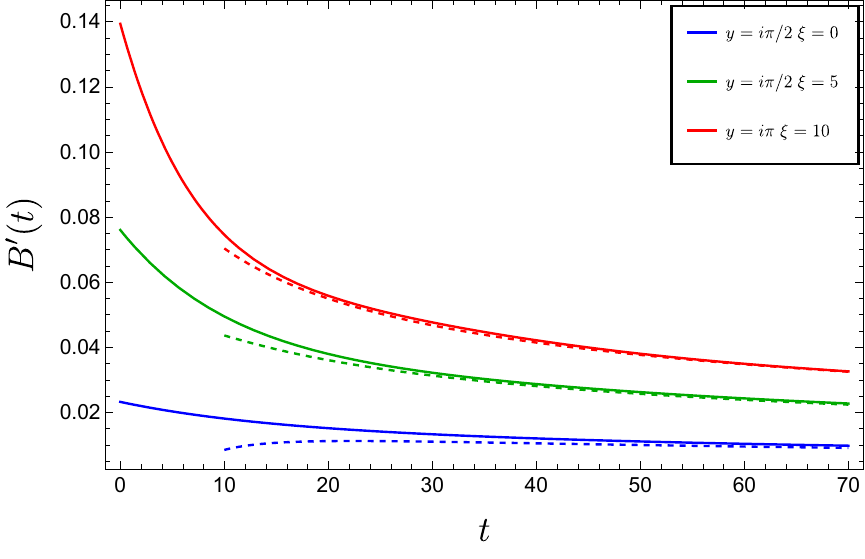}
\caption{}
\label{fig:octCMEuclidean}
\end{subfigure}
\caption{(a) $B'(t)$ for Lorentzian kinematic points. Asymptotic expansions of $B'(t)$ are given by the dashed graphs.   (b)  $B'(t)$ for Euclidean kinematic points. Asymptotic expansions of $B'(t)$ are given by the dashed graphs.
} 
\label{fig:octCM}
\end{figure}
\subsubsection{Octagon correlation function}
We will study complete monotonicity of $-\log D_0(y,\xi)$, which has previously been shown not to be Stieltjes for generic kinematics. The strong-coupling expansion is known to be \cite{Belitsky_2020}
\begin{align}
    -\log D_0 &\sim A_0 \sqrt{\lambda}-A_1\log\lambda + \dots\\
    A_0 &= \frac{1}{4 \pi^2} \int_0^\infty dx \, x \partial_x \log \left( \frac{\cosh \sqrt{x^2+\xi^2}-\cosh \xi}{\cosh \sqrt{x^2+\xi^2}+\cosh y}\right)\\
    A_1 &= \frac{1}{4}.
\end{align}
Together with equation \eqref{eq:tauberianobs} this gives the large $t$ asymptotics for $B'(t)$.

We test complete monotonicity for various kinematical points. For Lorentzian kinematics we choose $(y,\xi)=(0,0),\, (5,0),\, (0,5), \, (5,5)$. For Euclidean kinematics we choose  $(y,\xi)=(i \pi/2,0),\, (i\pi/2,5),\, (i \pi,5)$. All of these points except $(0,0),\, (5,0),\, (i \pi/2,0)$ violate the Hankel matrix constraints considered in Section~\ref{sec:failuresoctagon}. The results are shown in Figure~\ref{fig:octCM}. For all kinematical points, the perturbatively computed $B'(t)$ and the asymptotic estimate obtained from \eqref{eq:tauberianobs} are positive thereby establishing positivity for all $t>0$. 

Hence, complete monotonicity is satisfied at kinematical points where the Stieltjes property is not.  We believe that complete monotonicity holds for generic kinematical points.

\section{Strong-coupling expansion of the circular Wilson loop from the dispersive representation}
\label{app:circular}

In this Appendix, we derive the strong-coupling expansion of the logarithm of the circular Wilson loop from its Mellin--Barnes representation. Using \eqref{eq:logW-Stieltjes} and \eqref{eq:mb-master}, we obtain the representation
\begin{equation}\label{eq:WcircMellin}
    \log W = \int_{\delta-i\infty}^{\delta+i\infty} \frac{dj}{2\pi i} \, \frac{\pi}{\sin \pi j} \lambda^j \hat{\nu}(j), \qquad \hat{\nu}(j)=\int_0^{1/j_{1,1}^2}dt\, t^{j-1} \sum_{k=1}^\infty \theta\left(\frac{1}{j_{1,k}^2}-t \right).
\end{equation}
Here, $1/2<\operatorname{Re}\delta<1$. The analytic continuation of $\hat{\nu}(j)$ is more involved compared to the octagon anomalous dimension discussed in Subsection~\ref{sec:mb-octagon}. Doing the integration in the Mellin transform yields
\begin{equation}
    \hat{\nu}(j) = \frac{1}{j} \sum_{k=1}^\infty (j_{1,k})^{-2  j}.
\end{equation} 
There is no known analytic continuation of this sum. However, we can approximate $j_{1,k}$ by its asymptotic expansion for large $k$ and then analytically continue the result. For this we use the expansion \cite{McMahon:1895xs}
\begin{equation}
     j_{1,k}\sim \pi(k+\frac{1}{4})-\frac{3}{8} \frac{1}{\pi(k+\frac{1}{4})}+\mathcal{O}\left(\left(k+\tfrac{1}{4}\right)^{-2}\right).
\end{equation}
Plugging this into the $\hat{\nu}(j)$ yields an approximate function $ \hat{\nu}^{\mathrm{approx}}(j)$:
\begin{align} \label{eq:asymptoticmellincirc}
    \hat{\nu}^{\mathrm{approx}}(j) &= \frac{1}{j} \sum_{k=1}^\infty \pi^{-2j}(k+\frac{1}{4})^{-2j}\bigg(1-\frac{3}{8}\frac{1}{\pi^2(k+1/4)^2} \bigg)^{-2j}\\
    &=  \frac{1}{j}\sum_{m=0}^\infty \binom{m+2j-1}{m}\pi^{-2j-2m}\bigg(\frac{3}{8}\bigg)^m \zeta\left(2j+2m, \frac{5}{4}\right).
\end{align}
In going from the first to the second line, we expanded $(1-\tfrac{3}{8}\tfrac{1}{\pi^2(k+1/4)^2})^{-2j}$ into an infinite series, and introduced the Hurwitz zeta function $\zeta(s,q) = \sum_{k=0}^\infty \tfrac{1}{(k+q)^s}$. Now, \eqref{eq:asymptoticmellincirc} is in a form that can be analytically continued to $j\in \mathbb{C}$. The Mellin transform has simple poles at $j=0$ from the $1/j$ and from the Hurwitz zeta functions whenever $2j+2m =1$, $m\in \mathbb{N}_0$. All other pieces in \eqref{eq:asymptoticmellincirc} are analytic in the entire complex plane. We can plug \eqref{eq:asymptoticmellincirc} back into the Mellin--Barnes representation \eqref{eq:WcircMellin}
\begin{align}
    \log W^{\mathrm{approx}}(\lambda) \approx \int_{\delta-i\infty}^{\delta+i \infty} \frac{dj}{2\pi i}\; \frac{\pi}{\sin \pi j} \frac{\lambda^j}{j} \sum_{m=0}^\infty \binom{m+2j-1}{m}\pi^{-2j-2m}\bigg(\frac{3}{8}\bigg)^m \zeta\left(2j+2m, \frac{5}{4}\right).
\end{align}
There are poles at $j=1/2,0,-1/2,-3/2, \dots$ coming from the Mellin transform and at $j=0,1,2,\dots$ from the $\pi/\sin \pi j$. In contrast to the octagon anomalous dimension in \eqref{eq:OctagonMellinPert}, there is no cancellation of poles.  Computing the first terms in the strong-coupling expansion yields
\begin{equation} \label{eq:Wasymptoticstrong}
    \log W (\lambda)^{\mathrm{approx}} \sim \sqrt{\lambda} -\frac{3}{4}\log \lambda +\frac{3}{2}\log \pi + 8 \zeta'(0,\frac{5}{4})+\sum_{m=1}^\infty 2^{1-3m}3^m \pi^{-2m} \frac{\zeta(2m,\frac{5}{4})}{m} + \dots . 
\end{equation}
This can be compared to the exact result of the strong-coupling expansion given in \cite{bajnok2024solvingfourdimensionalsuperconformalyangmills} 
\begin{equation}\label{eq:CircStrong}
    \log W(\lambda) \sim \sqrt{\lambda}-\frac{3}{4}\log \lambda-\frac{1}{2}\log \frac{\pi}{2}-\frac{3}{8\sqrt{\lambda}}-\frac{3}{16\lambda}-\frac{21}{128\lambda^{3/2}}-\dots.
\end{equation}
Indeed, the first two terms of the asymptotic expansion match the exact result. We can compute higher corrections to \eqref{eq:Wasymptoticstrong} and compare the numerical values
\begin{align}
    \log W(\lambda)^{\mathrm{approx}} &\sim \sqrt{\lambda}- \frac{3}{4}\log \lambda - 0.2256- \frac{3}{8\sqrt{\lambda}}-\frac{0.191}{\lambda}-\frac{0.14}{\lambda^{3/2}} -\dots\\
    \log W(\lambda) &\sim \sqrt{\lambda}-\frac{3}{4}\log \lambda-0.2258-\frac{3}{8\sqrt{\lambda}}-\frac{0.188}{\lambda}-\frac{0.16}{\lambda^{3/2}}-\dots.
\end{align}
Here, we left the factors that match exactly as fractions and evaluated the remaining pieces numerically. It is very interesting that the analytic continuation using the asymptotic expansion of $j_{1,k}$ captures some of the strong-coupling expansion coefficients exactly. 
\bibliographystyle{JHEP}
\bibliography{refs}

@book{Weinberg:1995mt,
    author = "Weinberg, Steven",
    title = "{The Quantum theory of fields. Vol. 1: Foundations}",
    doi = "10.1017/CBO9781139644167",
    isbn = "978-0-521-67053-1, 978-0-511-25204-4",
    publisher = "Cambridge University Press",
    year = "2005"
}

@book{Eden:1966dnq,
    author = "Eden, Richard John and Landshoff, Peter V. and Olive, David I. and Polkinghorne, John Charlton",
    title = "{The analytic S-matrix}",
    isbn = "978-0-521-04869-9",
    publisher = "Cambridge Univ. Press",
    year = "1966"
}

@article{Colangelo:2000dp,
    author = "Colangelo, Pietro and Khodjamirian, Alexander",
    editor = "Shifman, M. and Ioffe, Boris",
    title = "{QCD sum rules, a modern perspective}",
    eprint = "hep-ph/0010175",
    archivePrefix = "arXiv",
    doi = "10.1142/9789812810458_0033",
    pages = "1495--1576",
    year = "2000"
}

@article{Arkani-Hamed:2020blm,
    author = "Arkani-Hamed, Nima and Huang, Tzu-Chen and Huang, Yu-tin",
    title = "{The EFT-Hedron}",
    eprint = "2012.15849",
    archivePrefix = "arXiv",
    doi = "10.1007/JHEP05(2021)259",
    journal = "JHEP",
    volume = "05",
    pages = "259",
    year = "2021"
}

@article{Poland:2018epd,
    author = "Poland, David and Rychkov, Slava and Vichi, Alessandro",
    title = "{The Conformal Bootstrap: Theory, Numerical Techniques, and Applications}",
    eprint = "1805.04405",
    archivePrefix = "arXiv",
    doi = "10.1103/RevModPhys.91.015002",
    journal = "Rev. Mod. Phys.",
    volume = "91",
    pages = "015002",
    year = "2019"
}

@article{Bellazzini_2021,
    author = "Bellazzini, Brando and Elias Mir{\'o}, Joan and Rattazzi, Riccardo and Riembau, Marc and Riva, Francesco",
    title = "{Positive moments for scattering amplitudes}",
    eprint = "2011.00037",
    archivePrefix = "arXiv",
    doi = "10.1103/PhysRevD.104.036006",
    journal = "Phys. Rev. D",
    volume = "104",
    number = "3",
    pages = "036006",
    year = "2021"
}

@article{henn2025positivitypropertiesscatteringamplitudes,
    author = "Henn, Johannes and Raman, Prashanth",
    title = "{Positivity properties of scattering amplitudes}",
    eprint = "2407.05755",
    archivePrefix = "arXiv",
    doi = "10.1007/JHEP04(2025)150",
    journal = "JHEP",
    volume = "04",
    pages = "150",
    year = "2025"
}

@article{ditsch2026approximatingfeynmanintegralsusing,
    author = "Ditsch, Sara and Henn, Johannes M. and Raman, Prashanth",
    title = "{Approximating Feynman integrals using complete monotonicity and Stieltjes properties}",
    eprint = "2512.18499",
    archivePrefix = "arXiv",
    primaryClass = "hep-th",
    reportNumber = "MPP-2025-226, TUM-HEP-1584/25",
    doi = "10.1007/JHEP05(2026)122",
    journal = "JHEP",
    volume = "05",
    pages = "122",
    year = "2026"
}

@article{PhysRev.184.1231,
    author = "Bender, Carl M. and Wu, Tai Tsun",
    title = "{Anharmonic oscillator}",
    doi = "10.1103/PhysRev.184.1231",
    journal = "Phys. Rev.",
    volume = "184",
    pages = "1231--1260",
    year = "1969"
}

@article{SIMON197076,
    author = "Simon, B.",
    title = "{Coupling constant analyticity for the anharmonic oscillator}",
    doi = "10.1016/0003-4916(70)90240-X",
    journal = "Annals Phys.",
    volume = "58",
    pages = "76--136",
    year = "1970"
}

@book{Bender:1999box,
    author = "Bender, Carl M. and Orszag, Steven A.",
    title = "{Advanced Mathematical Methods for Scientists and Engineers I}",
    doi = "10.1007/978-1-4757-3069-2",
    publisher = "Springer",
    year = "1999"
}

@article{Dorigoni_2019,
    author = "Dorigoni, Daniele",
    title = "{An Introduction to Resurgence, Trans-Series and Alien Calculus}",
    eprint = "1411.3585",
    archivePrefix = "arXiv",
    doi = "10.1016/j.aop.2019.167914",
    journal = "Annals Phys.",
    volume = "409",
    pages = "167914",
    year = "2019"
}

@book{WidderWidder+2015,
    title = "{Laplace Transform}",
    author = "David Vernon Widder",
    publisher = "Princeton University Press",
    doi = "doi:10.1515/9781400876457",
    year = "2015"
}

@article{merkle2012completelymonotonefunctions,
    title = "{Completely monotone functions - a digest}",
    author = "Milan Merkle",
    year = "2012",
    eprint = "1211.0900",
    archivePrefix = "arXiv"
}

@book{Baker_Graves-Morris_1996,
    place = {Cambridge},
    edition = {2},
    series = {Encyclopedia of Mathematics and its Applications},
    title = {Pad\'e Approximants},
    publisher = {Cambridge University Press},
    author = {Baker, George A. and Graves-Morris, Peter},
    year = {1996}
}

@article{schmudgen2020lecturesmomentproblem,
      title={Ten Lectures on the Moment Problem}, 
      author={Konrad Schm\"{u}dgen},
      year={2020},
      eprint={2008.12698},
      archivePrefix={arXiv},
      primaryClass={math.FA},
      url={https://arxiv.org/abs/2008.12698}, 
}

@inbook{doi:10.1137/1.9781611976397.ch3,
    title = {Function theoretic methods in the moment problem},
    booktitle = {The Classical Moment Problem and Some Related Questions in Analysis},
    author={Naum Ilyich Akhiezer},
    pages = {90--137},
    publisher = {Society for Industrial and Applied Mathematics},
    doi = {10.1137/1.9781611976397.ch3},
    year = {2020}
}

@article{Basdevant:1972fe,
    author = "Basdevant, J. L.",
    title = "{The Pad\'{e} approximation and its physical applications}",
    doi = "10.1002/prop.19720200502",
    journal = "Fortsch. Phys.",
    volume = "20",
    pages = "283--331",
    year = "1972"
}

@book{feller1,
    author = {Feller, William},
    title = {An Introduction to Probability Theory and Its Applications},
    publisher = {Wiley},
    volume = {1},
    year = {1968}
}

@article{Maldacena:1997re,
    author = "Maldacena, Juan Martin",
    title = "{The Large $N$ limit of superconformal field theories and supergravity}",
    eprint = "hep-th/9711200",
    archivePrefix = "arXiv",
    doi = "10.4310/ATMP.1998.v2.n2.a1",
    journal = "Adv. Theor. Math. Phys.",
    volume = "2",
    pages = "231--252",
    year = "1998"
}

@article{Beisert:2006ez,
    author = "Beisert, Niklas and Eden, Burkhard and Staudacher, Matthias",
    title = "{Transcendentality and Crossing}",
    eprint = "hep-th/0610251",
    archivePrefix = "arXiv",
    doi = "10.1088/1742-5468/2007/01/P01021",
    journal = "J. Stat. Mech.",
    volume = "0701",
    pages = "P01021",
    year = "2007"
}

@article{Beisert:2010jr,
    author = "Beisert, Niklas and others",
    title = "{Review of AdS/CFT Integrability: An Overview}",
    eprint = "1012.3982",
    archivePrefix = "arXiv",
    doi = "10.1007/s11005-011-0529-2",
    journal = "Lett. Math. Phys.",
    volume = "99",
    pages = "3--32",
    year = "2012"
}

@article{Pestun_2012,
    author = "Pestun, Vasily",
    title = "{Localization of gauge theory on a four-sphere and supersymmetric Wilson loops}",
    eprint = "0712.2824",
    archivePrefix = "arXiv",
    doi = "10.1007/s00220-012-1485-0",
    journal = "Commun. Math. Phys.",
    volume = "313",
    pages = "71--129",
    year = "2012"
}

@article{Polyakov:1980ca,
    author = "Polyakov, Alexander M.",
    title = "{Gauge Fields as Rings of Glue}",
    doi = "10.1016/0550-3213(80)90507-6",
    journal = "Nucl. Phys. B",
    volume = "164",
    pages = "171--188",
    year = "1980"
}

@article{Korchemsky:1987wg,
    author = "Korchemsky, G. P. and Radyushkin, A. V.",
    title = "{Renormalization of the Wilson Loops Beyond the Leading Order}",
    doi = "10.1016/0550-3213(87)90277-X",
    journal = "Nucl. Phys. B",
    volume = "283",
    pages = "342--364",
    year = "1987"
}

@article{KORCHEMSKAYA1992169,
    author = "Korchemskaya, I. A. and Korchemsky, G. P.",
    title = "{On lightlike Wilson loops}",
    doi = "10.1016/0370-2693(92)91895-G",
    journal = "Phys. Lett. B",
    volume = "287",
    pages = "169--175",
    year = "1992"
}

@article{collins2004factorizationhardprocessesqcd,
    author = "Collins, John C. and Soper, Davison E. and Sterman, George F.",
    title = "{Factorization of Hard Processes in QCD}",
    eprint = "hep-ph/0409313",
    archivePrefix = "arXiv",
    doi = "10.1142/9789814503266_0001",
    journal = "Adv. Ser. Direct. High Energy Phys.",
    volume = "5",
    pages = "1--91",
    year = "1989"
}

@article{Bern_2007,
    author = "Bern, Zvi and Czakon, Michael and Dixon, Lance J. and Kosower, David A. and Smirnov, Vladimir A.",
    title = "{The Four-Loop Planar Amplitude and Cusp Anomalous Dimension in Maximally Supersymmetric Yang-Mills Theory}",
    eprint = "hep-th/0610248",
    archivePrefix = "arXiv",
    doi = "10.1103/PhysRevD.75.085010",
    journal = "Phys. Rev. D",
    volume = "75",
    pages = "085010",
    year = "2007"
}

@article{Basso_2008,
    author = "Basso, B. and Korchemsky, G. P. and Kotanski, J.",
    title = "{Cusp anomalous dimension in maximally supersymmetric Yang-Mills theory at strong coupling}",
    eprint = "0708.3933",
    archivePrefix = "arXiv",
    doi = "10.1103/PhysRevLett.100.091601",
    journal = "Phys. Rev. Lett.",
    volume = "100",
    pages = "091601",
    year = "2008"
}

@article{Basso_2020,
    author = "Basso, Benjamin and Dixon, Lance J. and Papathanasiou, Georgios",
    title = "{Origin of the Six-Gluon Amplitude in Planar $N=4$ Supersymmetric Yang-Mills Theory}",
    eprint = "2001.05460",
    archivePrefix = "arXiv",
    doi = "10.1103/PhysRevLett.124.161603",
    journal = "Phys. Rev. Lett.",
    volume = "124",
    pages = "161603",
    year = "2020"
}

@article{Dorigoni_2015,
    author = "Dorigoni, Daniele and Hatsuda, Yasuyuki",
    title = "{Resurgence of the Cusp Anomalous Dimension}",
    eprint = "1506.03763",
    archivePrefix = "arXiv",
    doi = "10.1007/JHEP09(2015)138",
    journal = "JHEP",
    volume = "09",
    pages = "138",
    year = "2015"
}

@article{Correa_2012,
    author = "Correa, Diego and Henn, Johannes and Maldacena, Juan and Sever, Amit",
    title = "{An exact formula for the radiation of a moving quark in N=4 super Yang Mills}",
    eprint = "1202.4455",
    archivePrefix = "arXiv",
    doi = "10.1007/JHEP06(2012)048",
    journal = "JHEP",
    volume = "06",
    pages = "048",
    year = "2012"
}

@article{Drukker_2001,
    author = "Drukker, Nadav and Gross, David J.",
    title = "{An Exact prediction of N=4 SUSYM theory for string theory}",
    eprint = "hep-th/0010274",
    archivePrefix = "arXiv",
    doi = "10.1063/1.1372177",
    journal = "J. Math. Phys.",
    volume = "42",
    pages = "2896--2914",
    year = "2001"
}

@article{Erickson_2000,
    author = "Erickson, J. K. and Semenoff, G. W. and Zarembo, K.",
    title = "{Wilson loops in N=4 supersymmetric Yang-Mills theory}",
    eprint = "hep-th/0003055",
    archivePrefix = "arXiv",
    doi = "10.1016/S0550-3213(00)00300-X",
    journal = "Nucl. Phys. B",
    volume = "582",
    pages = "155--175",
    year = "2000"
}

@article{Chicherin_2016,
    author = "Chicherin, Dmitry and Drummond, James and Heslop, Paul and Sokatchev, Emery",
    title = "{All three-loop four-point correlators of half-BPS operators in planar $ \mathcal{N} $ = 4 SYM}",
    eprint = "1512.02926",
    archivePrefix = "arXiv",
    doi = "10.1007/JHEP08(2016)053",
    journal = "JHEP",
    volume = "08",
    pages = "053",
    year = "2016"
}

@article{Coronado_2020,
    author = "Coronado, Frank",
    title = "{Bootstrapping the Simplest Correlator in Planar $\mathcal N = 4$ Supersymmetric Yang-Mills Theory to All Loops}",
    eprint = "1811.03282",
    archivePrefix = "arXiv",
    doi = "10.1103/PhysRevLett.124.171601",
    journal = "Phys. Rev. Lett.",
    volume = "124",
    pages = "171601",
    year = "2020"
}

@article{Belitsky_2020,
    author = "Belitsky, A. V. and Korchemsky, G. P.",
    title = "{Octagon at finite coupling}",
    eprint = "2003.01121",
    archivePrefix = "arXiv",
    doi = "10.1007/JHEP07(2020)219",
    journal = "JHEP",
    volume = "07",
    pages = "219",
    year = "2020"
}

@article{Belitsky_2021,
    author = "Belitsky, A. V. and Korchemsky, G. P.",
    title = "{Crossing bridges with strong Szeg{\H{o}} limit theorem}",
    eprint = "2006.01831",
    archivePrefix = "arXiv",
    doi = "10.1007/JHEP04(2021)257",
    journal = "JHEP",
    volume = "04",
    pages = "257",
    year = "2021"
}

@article{Bajnok:2024epf,
    author = "Bajnok, Zoltan and Boldis, Bercel and Korchemsky, Gregory P.",
    title = "{Tracy-Widom Distribution in Four-Dimensional Supersymmetric Yang-Mills Theories}",
    eprint = "2403.13050",
    archivePrefix = "arXiv",
    doi = "10.1103/PhysRevLett.133.031601",
    journal = "Phys. Rev. Lett.",
    volume = "133",
    pages = "031601",
    year = "2024"
}

@article{Bajnok:2024bqr,
    author = "Bajnok, Zoltan and Boldis, Bercel and Korchemsky, Gregory P.",
    title = "{Exploring superconformal Yang-Mills theories through matrix Bessel kernels}",
    eprint = "2412.08732",
    archivePrefix = "arXiv",
    doi = "10.21468/SciPostPhys.19.1.004",
    journal = "SciPost Phys.",
    volume = "19",
    pages = "004",
    year = "2025"
}

@article{bajnok2024solvingfourdimensionalsuperconformalyangmills,
    author = "Bajnok, Zoltan and Boldis, Bercel and Korchemsky, Gregory P.",
    title = "{Solving four-dimensional superconformal Yang-Mills theories with Tracy-Widom distribution}",
    eprint = "2409.17227",
    archivePrefix = "arXiv",
    doi = "10.1007/JHEP04(2025)005",
    journal = "JHEP",
    volume = "04",
    pages = "005",
    year = "2025"
}

@article{bajnok2025universalityresurgencegeneralizedtracywidom,
    author = "Bajnok, Zoltan and Boldis, Bercel and le Plat, Dennis",
    title = "{Universality in the resurgence of generalized Tracy-Widom distributions}",
    eprint = "2509.20302",
    archivePrefix = "arXiv",
    doi = "10.1016/j.physletb.2026.140232",
    journal = "Phys. Lett. B",
    volume = "874",
    pages = "140232",
    year = "2026"
}

@article{McMahon:1895xs,
    author = "McMahon, James",
    title = "{On the roots of the Bessel and certain related functions}",
    doi = "10.2307/1967501",
    journal = "Annals Math.",
    volume = "9",
    pages = "23--30",
    year = "1895"
}

@article{Marboe_2015,
    author = "Marboe, Christian and Volin, Dmytro",
    title = "{Quantum spectral curve as a tool for a perturbative quantum field theory}",
    eprint = "1411.4758",
    archivePrefix = "arXiv",
    doi = "10.1016/j.nuclphysb.2015.08.021",
    journal = "Nucl. Phys. B",
    volume = "899",
    pages = "810--847",
    year = "2015"
}

@article{Arkani_Hamed_2022,
    author = "Arkani-Hamed, Nima and Henn, Johannes and Trnka, Jaroslav",
    title = "{Nonperturbative negative geometries: amplitudes at strong coupling and the amplituhedron}",
    eprint = "2112.06956",
    archivePrefix = "arXiv",
    doi = "10.1007/JHEP03(2022)108",
    journal = "JHEP",
    volume = "03",
    pages = "108",
    year = "2022"
}

@article{Raman_2021,
    author = "Raman, Prashanth and Sinha, Aninda",
    title = "{QFT, EFT and GFT}",
    eprint = "2107.06559",
    archivePrefix = "arXiv",
    doi = "10.1007/JHEP12(2021)203",
    journal = "JHEP",
    volume = "12",
    pages = "203",
    year = "2021"
}

@article{Belitsky:2019fan,
    author = "Belitsky, A. V. and Korchemsky, G. P.",
    title = "{Exact null octagon}",
    eprint = "1907.13131",
    archivePrefix = "arXiv",
    primaryClass = "hep-th",
    reportNumber = "IPhT-T19/097",
    doi = "10.1007/JHEP05(2020)070",
    journal = "JHEP",
    volume = "05",
    pages = "070",
    year = "2020"
}

@article{Magnea:1990zb,
    author = "Magnea, Lorenzo and Sterman, George F.",
    title = "{Analytic continuation of the Sudakov form-factor in QCD}",
    reportNumber = "ITP-SB-90-43",
    doi = "10.1103/PhysRevD.42.4222",
    journal = "Phys. Rev. D",
    volume = "42",
    pages = "4222--4227",
    year = "1990"
}

@article{Benna:2006nd,
    author = "Benna, M. K. and Benvenuti, S. and Klebanov, I. R. and Scardicchio, A.",
    title = "{A Test of the AdS/CFT correspondence using high-spin operators}",
    eprint = "hep-th/0611135",
    archivePrefix = "arXiv",
    doi = "10.1103/PhysRevLett.98.131603",
    journal = "Phys. Rev. Lett.",
    volume = "98",
    pages = "131603",
    year = "2007"
}

@article{Kodaira:1981nh,
    author = "Kodaira, Jiro and Trentadue, Luca",
    title = "{Summing Soft Emission in QCD}",
    reportNumber = "SLAC-PUB-2807",
    doi = "10.1016/0370-2693(82)90907-8",
    journal = "Phys. Lett. B",
    volume = "112",
    pages = "66",
    year = "1982"
}

@article{Bender:2001jq,
    author = "Bender, Carl M. and Weniger, E.",
    title = "{Numerical evidence that the perturbation expansion for a nonHermitian Hamiltonian is Stieltjes}",
    eprint = "math-ph/0010007",
    archivePrefix = "arXiv",
    reportNumber = "WUHEP-00-18",
    doi = "10.1063/1.1362287",
    journal = "J. Math. Phys.",
    volume = "42",
    pages = "2167--2183",
    year = "2001"
}

@article{Grecchi_2009,
doi = {10.1088/1751-8113/42/42/425208},
url = {https://doi.org/10.1088/1751-8113/42/42/425208},
year = {2009},
month = {oct},
publisher = {},
volume = {42},
number = {42},
pages = {425208},
author = {Grecchi, Vincenzo and Maioli, Marco and Martinez, André},
title = {Padé summability of the cubic oscillator},
journal = {Journal of Physics A: Mathematical and Theoretical}
}

@article{Graffi:1970erh,
    author = "Graffi, S. and Grecchi, V. and Simon, B.",
    title = "{Borel summability: Application to the anharmonic oscillator}",
    doi = "10.1016/0370-2693(70)90564-2",
    journal = "Phys. Lett. B",
    volume = "32",
    pages = "631--634",
    year = "1970"
}

@article{Alday:2007mf,
    author = "Alday, Luis F. and Maldacena, Juan Martin",
    title = "{Comments on operators with large spin}",
    eprint = "0708.0672",
    archivePrefix = "arXiv",
    primaryClass = "hep-th",
    doi = "10.1088/1126-6708/2007/11/019",
    journal = "JHEP",
    volume = "11",
    pages = "019",
    year = "2007"
}

@article{Aniceto:2015rua,
    author = "Aniceto, In{\^e}s",
    title = "{The Resurgence of the Cusp Anomalous Dimension}",
    eprint = "1506.03388",
    archivePrefix = "arXiv",
    primaryClass = "hep-th",
    doi = "10.1088/1751-8113/49/6/065403",
    journal = "J. Phys. A",
    volume = "49",
    pages = "065403",
    year = "2016"
}

@article{Basso:2022ruw,
    author = "Basso, Benjamin and Dixon, Lance J. and Liu, Yu-Ting and Papathanasiou, Georgios",
    title = "{All-Orders Quadratic-Logarithmic Behavior for Amplitudes}",
    eprint = "2211.12555",
    archivePrefix = "arXiv",
    primaryClass = "hep-th",
    reportNumber = "SLAC--PUB--17710, DESY-22-182",
    doi = "10.1103/PhysRevLett.130.111602",
    journal = "Phys. Rev. Lett.",
    volume = "130",
    number = "11",
    pages = "111602",
    year = "2023"
}

@article{Kostov:2019auq,
    author = "Kostov, Ivan and Petkova, Valentina B. and Serban, Didina",
    title = "{The Octagon as a Determinant}",
    eprint = "1905.11467",
    archivePrefix = "arXiv",
    primaryClass = "hep-th",
    doi = "10.1007/JHEP11(2019)178",
    journal = "JHEP",
    volume = "11",
    pages = "178",
    year = "2019"
}

@article{Caron-Huot:2021usw,
    author = "Caron-Huot, Simon and Coronado, Frank",
    title = "{Ten dimensional symmetry of $ \mathcal{N} $ = 4 SYM correlators}",
    eprint = "2106.03892",
    archivePrefix = "arXiv",
    primaryClass = "hep-th",
    doi = "10.1007/JHEP03(2022)151",
    journal = "JHEP",
    volume = "03",
    pages = "151",
    year = "2022"
}

@article{Aniceto:2018bis,
    author = "Aniceto, In{\^e}s and Basar, Gokce and Schiappa, Ricardo",
    title = "{A Primer on Resurgent Transseries and Their Asymptotics}",
    eprint = "1802.10441",
    archivePrefix = "arXiv",
    primaryClass = "hep-th",
    reportNumber = "NSF-ITP-17-153",
    doi = "10.1016/j.physrep.2019.02.003",
    journal = "Phys. Rept.",
    volume = "809",
    pages = "1--135",
    year = "2019"
}

@article{Korchemsky:2025mla,
    author = "Korchemsky, G. P.",
    title = "{Lattice path combinatorics in superconformal Yang-Mills theories}",
    eprint = "2508.20901",
    archivePrefix = "arXiv",
    primaryClass = "hep-th",
    month = "8",
    year = "2025"
}

@article{Beccaria:2022ypy,
    author = "Beccaria, M. and Korchemsky, G. P. and Tseytlin, A. A.",
    title = "{Strong coupling expansion in N = 2 superconformal theories and the Bessel kernel}",
    eprint = "2207.11475",
    archivePrefix = "arXiv",
    primaryClass = "hep-th",
    reportNumber = "IPhT--T22/040, Imperial-TP-AT-2022-04",
    doi = "10.1007/JHEP09(2022)226",
    journal = "JHEP",
    volume = "09",
    pages = "226",
    year = "2022"
}

@article{Adams:2006sv,
    author = "Adams, Allan and Arkani-Hamed, Nima and Dubovsky, Sergei and Nicolis, Alberto and Rattazzi, Riccardo",
    title = "{Causality, analyticity and an IR obstruction to UV completion}",
    eprint = "hep-th/0602178",
    archivePrefix = "arXiv",
    reportNumber = "CERN-PH-TH-2006-033, HUTP-06-A0005",
    doi = "10.1088/1126-6708/2006/10/014",
    journal = "JHEP",
    volume = "10",
    pages = "014",
    year = "2006"
}

@article{Minahan:2002ve,
    author = "Minahan, J. A. and Zarembo, K.",
    title = "{The Bethe ansatz for N=4 superYang-Mills}",
    eprint = "hep-th/0212208",
    archivePrefix = "arXiv",
    reportNumber = "UUITP-17-02, ITEP-TH-73-02",
    doi = "10.1088/1126-6708/2003/03/013",
    journal = "JHEP",
    volume = "03",
    pages = "013",
    year = "2003"
}

@article{Ambjorn:2005wa,
    author = "Ambjorn, Jan and Janik, Romuald A. and Kristjansen, Charlotte",
    title = "{Wrapping interactions and a new source of corrections to the spin-chain/string duality}",
    eprint = "hep-th/0510171",
    archivePrefix = "arXiv",
    reportNumber = "NORDITA-2005-67",
    doi = "10.1016/j.nuclphysb.2005.12.007",
    journal = "Nucl. Phys. B",
    volume = "736",
    pages = "288--301",
    year = "2006"
}

@article{Bajnok:2008bm,
    author = "Bajnok, Zoltan and Janik, Romuald A.",
    title = "{Four-loop perturbative Konishi from strings and finite size effects for multiparticle states}",
    eprint = "0807.0399",
    archivePrefix = "arXiv",
    primaryClass = "hep-th",
    doi = "10.1016/j.nuclphysb.2008.08.020",
    journal = "Nucl. Phys. B",
    volume = "807",
    pages = "625--650",
    year = "2009"
}

@article{Berenstein:2002jq,
    author = "Berenstein, David Eliecer and Maldacena, Juan Martin and Nastase, Horatiu Stefan",
    title = "{Strings in flat space and pp waves from N=4 superYang-Mills}",
    eprint = "hep-th/0202021",
    archivePrefix = "arXiv",
    doi = "10.1088/1126-6708/2002/04/013",
    journal = "JHEP",
    volume = "04",
    pages = "013",
    year = "2002"
}

@article{Plefka:2005bk,
    author = "Plefka, Jan",
    title = "{Spinning strings and integrable spin chains in the AdS/CFT correspondence}",
    eprint = "hep-th/0507136",
    archivePrefix = "arXiv",
    reportNumber = "AEI-2005-124",
    doi = "10.12942/lrr-2005-9",
    journal = "Living Rev. Rel.",
    volume = "8",
    pages = "9",
    year = "2005"
}

@article{Roiban:2011fe,
    author = "Roiban, R. and Tseytlin, A. A.",
    title = "{Semiclassical string computation of strong-coupling corrections to dimensions of operators in Konishi multiplet}",
    eprint = "1102.1209",
    archivePrefix = "arXiv",
    primaryClass = "hep-th",
    doi = "10.1016/j.nuclphysb.2011.02.016",
    journal = "Nucl. Phys. B",
    volume = "848",
    pages = "251--267",
    year = "2011"
}

@article{Beisert:2005fw,
    author = "Beisert, Niklas and Staudacher, Matthias",
    title = "{Long-range psu(2,2|4) Bethe Ansatze for gauge theory and strings}",
    eprint = "hep-th/0504190",
    archivePrefix = "arXiv",
    reportNumber = "AEI-2005-092, PUTP-2159",
    doi = "10.1016/j.nuclphysb.2005.06.038",
    journal = "Nucl. Phys. B",
    volume = "727",
    pages = "1--62",
    year = "2005"
}

@article{Korchemsky:1988si,
    author = "Korchemsky, G. P.",
    title = "{Asymptotics of the Altarelli-Parisi-Lipatov Evolution Kernels of Parton Distributions}",
    reportNumber = "JINR-E2-88-717",
    doi = "10.1142/S0217732389001453",
    journal = "Mod. Phys. Lett. A",
    volume = "4",
    pages = "1257--1276",
    year = "1989"
}

@article{Costin:2021bay,
    author = "Costin, Ovidiu and Dunne, Gerald V.",
    title = "{Conformal and uniformizing maps in Borel analysis}",
    eprint = "2108.01145",
    archivePrefix = "arXiv",
    primaryClass = "hep-th",
    doi = "10.1140/epjs/s11734-021-00267-x",
    journal = "Eur. Phys. J. ST",
    volume = "230",
    number = "12-13",
    pages = "2679--2690",
    year = "2021"
}

@article{Kruczenski:2022lot,
    author = "Kruczenski, Martin and Penedones, Joao and van Rees, Balt C.",
    title = "{Snowmass White Paper: S-matrix Bootstrap}",
    eprint = "2203.02421",
    archivePrefix = "arXiv",
    primaryClass = "hep-th",
    month = "3",
    year = "2022"
}

@article{Arkani-Hamed:2013jha,
    author = "Arkani-Hamed, Nima and Trnka, Jaroslav",
    title = "{The Amplituhedron}",
    eprint = "1312.2007",
    archivePrefix = "arXiv",
    primaryClass = "hep-th",
    doi = "10.1007/JHEP10(2014)030",
    journal = "JHEP",
    volume = "10",
    pages = "030",
    year = "2014"
}

@article{Dunne:2015eaa,
    author = {Dunne, Gerald V. and {\"U}nsal, Mithat},
    title = "{What is QFT? Resurgent trans-series, Lefschetz thimbles, and new exact saddles}",
    eprint = "1511.05977",
    archivePrefix = "arXiv",
    primaryClass = "hep-lat",
    doi = "10.22323/1.251.0010",
    journal = "PoS",
    volume = "LATTICE2015",
    pages = "010",
    year = "2016"
}

@article{Alday:2026cpk,
    author = {Alday, Luis F. and Armanini, Elisabetta and Belitsky, Andrei V. and H{\"a}ring, Kelian and Zhiboedov, Alexander},
    title = "{Walking Sudakov: From Cusp to Octagon}",
    eprint = "2605.16034",
    archivePrefix = "arXiv",
    primaryClass = "hep-th",
    reportNumber = "CERN-TH-2026-109",
    month = "5",
    year = "2026"
}

@article{Basso:2026vmx,
    author = "Basso, Benjamin and Fleury, Thiago and Kalu{\c{c}}, Erkan and Serban, Didina",
    title = "{Null limit of large-charge correlators in planar $\mathcal{N}=4$ Super-Yang-Mills theory}",
    eprint = "2606.24018",
    archivePrefix = "arXiv",
    primaryClass = "hep-th",
    month = "6",
    year = "2026"
}

\end{document}